\documentclass[twocolumn]{aastex63}

\received{}
\revised{}
\accepted{\today}
\submitjournal{AJ}

\usepackage{amsmath}
\usepackage{gensymb}    
\usepackage{xspace}
\usepackage{appendix}
\usepackage{mathtools}  
\usepackage{rotating}
\usepackage[encapsulated]{CJK}
\usepackage{ucs}
\usepackage[utf8x]{inputenc}
\usepackage[normalem]{ulem}

\newcommand{\umin}{$U_{\rm min}$\xspace}
\newcommand{\sigd}{$\Sigma_{\rm dust}$\xspace}
\newcommand{\qpah}{$q_{\rm \sc{PAH}}$\xspace}
\newcommand{\ostar}{$\Omega_*$\xspace}

\newcommand{\fpah}{$f_{\rm PAH}$\xspace}
\newcommand{\td}{$T_{\rm dust}$\xspace}
\newcommand{\lambdac}{$\lambda_{\rm c}$\xspace}

\newcommand{\ubar}{$\overline{U}$\xspace}
\newcommand{\avubar}{$\langle \overline{U} \rangle$\xspace}
\newcommand{\hi}{\ion{H}{1}~21~cm\xspace}
\newcommand{\coto}{${\rm CO~(2\to1)}$\xspace}

\defcitealias{DL07}{DL07}
\defcitealias{THEMIS}{THEMIS}
\defcitealias{Compiegne11}{MC11}
\defcitealias{Hensley+Draine_2020c}{HD21}
\newcommand{\dustbff}{\texttt{DustBFF}\xspace}

\newcommand{\BE}{broken-emissivity\xspace}
\newcommand{\SE}{simple-emissivity\xspace}

\newcommand{\spitzer}{{\it Spitzer}\xspace}
\newcommand{\herschel}{{\it Herschel}\xspace}
\newcommand{\dustem}{\texttt{DustEM}\xspace}

\newcommand{\rtf}{$\rm r_{25}$\xspace}
\newcommand{\logt}{${\rm log_{10}}$}
\newcommand{\msolpc}{M$_\odot$/pc$^2$\xspace}
\newcommand{\msol}{M$_\odot$\xspace}
\newcommand{\aco}{$\alpha_{\rm CO}$\xspace}

\shorttitle{Benchmarking Dust Emission Models}
\shortauthors{Chastenet, Sandstrom et al.}
\graphicspath{{./}{figures/}}

\begin{document}

\title{Benchmarking Dust Emission Models in M101}

\correspondingauthor{J\'er\'emy Chastenet}
\email{jchastenet@ucsd.edu}

\author[0000-0002-5235-5589]{J\'er\'emy Chastenet}
\affiliation{Center for Astrophysics and Space Sciences, Department of Physics, University of California, San Diego\\9500 Gilman Drive, La Jolla, CA 92093, USA}

\author[0000-0002-4378-8534]{Karin Sandstrom}
\affiliation{Center for Astrophysics and Space Sciences, Department of Physics, University of California, San Diego\\9500 Gilman Drive, La Jolla, CA 92093, USA}

\author[0000-0003-2551-7148]{I-Da Chiang \begin{CJK*}{UTF8}{bkai}(江宜達)\end{CJK*}}
\affiliation{Center for Astrophysics and Space Sciences, Department of Physics, University of California, San Diego\\9500 Gilman Drive, La Jolla, CA 92093, USA}

\author[0000-0001-7449-4638]{Brandon S. Hensley}
\affiliation{Princeton University Observatory, Peyton Hall, Princeton, NJ 08544-1001, USA}

\author[0000-0002-0846-936X]{Bruce T. Draine}
\affiliation{Princeton University Observatory, Peyton Hall, Princeton, NJ 08544-1001, USA}

\author[0000-0001-5340-6774]{Karl D. Gordon}
\affiliation{Space Telescope Science Institute, 3700 San Martin Drive, Baltimore, MD, 21218, USA}
\affiliation{Sterrenkundig Observatorium, Universiteit Gent, Gent, Belgium}

\author[0000-0001-9605-780X]{Eric W. Koch}
\affiliation{University of Alberta, Department of Physics, 4-183 CCIS, Edmonton AB T6G 2E1, Canada}

\author[0000-0002-2545-1700]{Adam K. Leroy}
\affiliation{Department of Astronomy, The Ohio State University, 4055 McPherson Laboratory, 140 West 18th Ave, Columbus, OH 43210, USA}

\author[0000-0003-4161-2639]{Dyas Utomo}
\affiliation{Department of Astronomy, The Ohio State University, 4055 McPherson Laboratory, 140 West 18th Ave, Columbus, OH 43210, USA}

\author[0000-0002-0012-2142]{Thomas~G.~Williams}
\affiliation{Max Planck Institut f{\"u}r Astronomie, K{\"o}nigstuhl 17, 69117 Heidelberg, Germany}

\begin{abstract}
We present a comparative study of four physical dust models and two single-temperature modified blackbody models by fitting them to the resolved WISE, \textit{Spitzer}, and \textit{Herschel} photometry of M101 (NGC~5457). Using identical data and a grid-based fitting technique, we compare the resulting dust and radiation field properties derived from the models. We find that the dust mass yielded by the different models can vary by up to factor of 3 (factor of 1.4 between physical models only), although the fits have similar quality. Despite differences in their definition of the carriers of the mid-IR aromatic features, all physical models show the same spatial variations for the abundance of that grain population.
Using the well determined metallicity gradient in M101 and resolved gas maps, we calculate an approximate upper limit on the dust mass as a function of radius. All physical dust models are found to exceed this maximum estimate over some range of galactocentric radii. We show that renormalizing the models to match the same Milky Way high latitude cirrus spectrum and abundance constraints can reduce the dust mass differences between models and bring the total dust mass below the maximum estimate at all radii. 
\end{abstract}


\section{Introduction}
Dust grains play key roles in processes that shape the interstellar medium (ISM) and galaxy evolution. They release photo-electrons that participate in heating gas \citep[e.g.][]{Wolfire95, WD01PE, Croxall12}, they shield dense molecular clouds from stellar UV radiation and aid their collapse \citep[e.g.][]{Fumagalli10, Byrne19}, they catalyze a number of chemical reactions, and offer surface area for the production of H$_2$ \citep[e.g.][see also a review by \citeauthor{Wakelam17}  \citeyear{Wakelam17}]{Bron14, Castellanos18, Thi20}. It is therefore critically important to understand dust properties and abundance, and how dust affects these processes.

One main way to infer dust properties in external galaxies is to interpret infrared (IR) spectral energy distributions (SEDs) with the aid of dust models. 
The near-to-mid-IR part of the spectrum is dominated by the emission of stochastically heated grains that do not achieve a steady-state equilibrium with the incident radiation field. At longer wavelengths, the emission is almost entirely due to grains in the thermal equilibrium, with a large enough radius to be constantly receiving and emitting photons. In this regime, the steady-state grain temperature is set by the strength of the incident radiation field.

Modified blackbody models are a convenient parametric representation of the emission from large grains in thermal equilibrium. As such, they provide good fits to the far-IR SED and yield satisfactory dust masses if correctly calibrated \citep[e.g.][]{Bianchi13}, since these grains contain most of the dust mass.
Large grains in thermal equilibrium are reasonably well described by a single temperature, in which case their emission can be represented by a blackbody radiation, $B_\nu$, modified by an effective grain opacity, $\kappa_\nu(\lambda)$, so that the grain emission $I_\nu \propto \kappa_\nu(\lambda)\ B_\nu(\lambda, T)$. The opacity as a function of wavelength is often described with a power-law with spectral index, $\beta$, such that $\kappa_\nu = \kappa_0 (\nu/\nu_0)^{\beta}$.
Several variations to the modified blackbody model have been used to fit the far-IR SED, for example multiple dust populations having different temperatures and $\beta$, or a different functional form for $\kappa_\nu$ such as a broken power-law. 

Physical dust models aim to reproduce the IR emission, extinction, and depletions, among other observations, with a self-consistent description of dust properties. Building these models requires specifying grain sizes, shapes, and chemical composition, which lead to the optical properties and heat capacities. These grain populations are then illuminated by a radiation field with a specified intensity and spectrum.
Once the radiation field is modeled, one can compare the predicted and observed dust emission. 
Dust extinction measured towards specific lines-of-sight (not high $A_V$) helps constrain the size distribution of grains, their composition, and total mass relative to H \citep[e.g.][]{Weingartner01}.
Depletion measurements provide important limits on the elemental abundance locked in dust grains \citep[e.g.][]{Jenkins09, Tchernyshyov15}.
Experimentally, many studies rely on material thought to be ISM dust analogs to provide laboratory constraints on optical properties and heat capacities \citep[e.g.][]{Richey13, Demyk17, Mutscke19}.

Both physical and modified blackbody models are almost always calibrated in the Milky Way (MW), where the the relevant observables of dust are well-constrained. This includes measurements of the diffuse emission and extinction per H column of the ISM at high Galactic latitudes also referred to as the Milky Way cirrus \citep[e.g.][]{Compiegne11, Guillet18}.
The cirrus is also a unique place where the interstellar radiation field that heats dust grains, a necessary component to constrain models, can be estimated.
The radiation field measured by \citet[][]{Mathis83} at the galactocentric distance $D_{\rm G}=10~$kpc is generally used to describe the starlight heating dust grains.

The stringency with which each model follows the elemental abundances locked in dust grains from depletion studies varies. Some models use them as strict constraints \citep[e.g.][]{Gordon14}. On the other hand, most physical models allow more flexibility in the mass of metals locked up in dust grains, to more closely match other, better constrained observables. For example the \citet[][]{Zubko04} dust models follow the depletion constraints strictly, while the \citet[][]{WD01Ext} can require up to $\sim 30$\% (assuming Si/H~=~36~ppm in the model) more silicon than observed in the cold neutral medium. However, the latter reproduce the observed extinction to a better degree than the former.

With the increasing number of observational constraints on dust models, the complexity of physical dust models has grown.
One of the earliest dust models described grains as a single mixture with a power-law size distribution \citep[][]{MRN77}. 
Later on, very small grains known as the Polycyclic Aromatic Hydrocarbons (PAHs) were suggested as responsible for the mid-IR emission features \citep[][]{Leger84, Allamandola85}, and included in dust models. While some dust models consider them to be defined by their size \citep[in the model description; e.g.][]{DL07, Compiegne11, Galliano11}, other models identify the mid-IR features carriers in the form of aromatic-rich mantles onto amorphous grains \citep[e.g.][and their hydrogenated amorphous hydrocarbons component]{Jones13}.

The precise nature of large grains is also uncertain.
The presence of amorphous silicate material in grains was demonstrated early on by conspicuous absorption features \citep[][]{MRN77, Kemper04} and included in dust models. But their exact composition remains an active research topic \citep[e.g.][]{Zeegers19}. While some models have a strong focus towards reproducing the observations \citep[e.g.][``astrosilicates'']{DL07}, other models are closely tied to new laboratory data \citep[e.g.][olivine and pyroxene]{Jones13}. 
Finally, with the growing amount of far-IR polarization data, new models emerge and take into account this important grain property \citep[e.g][]{Guillet18, Hensley+Draine_2020c}. 
Future missions and instruments are being developed to focus on polarization, and will bring new constraints to dust properties \citep[e.g. SOFIA/HAWC+:][]{Harper18}.

There are now several physical dust models available, that are all different in some---sometimes small---ways.
However, these small differences in modeling can lead to significant differences in the derived dust properties, as many studies have shown in nearby galaxies. For instance, \citet[][]{Gordon14} and \citet[][]{Chiang18} have used a number of blackbody variations (e.g. simple power-law emissivity, two temperatures, broken-emissivity) to model the far-IR SEDs of the Magellanic Clouds and M101, respectively. 
They both found that the dust mass derived by several blackbody variations (namely, simple power-law emissivity, two temperature) violates the available elemental content, making these approaches unlikely to be valid descriptions of dust grain emission.
In the Magellanic Clouds, \citet[][]{Chastenet17} found that dust masses can vary by almost an order of magnitude depending on the physical model chosen. In the Large Magellanic Cloud, \citet[][]{Paradis19} found that not all models require both neutral and ionized PAHs for a good fit at short wavelengths.

Other discrepancies may arise from simply using a different fitting approach, where uncertainties are treated differently, or with a different data-set. For instance, \citet[][]{Sandstrom10} found an increased abundance of PAHs in dense gas regions of the Small Magellanic Clouds. This behavior was confirmed, but with different PAH fractions in \citet{Chastenet19}, by using the same model with a different wavelength sampling.
Most of the uncertainties in comparing the results of the studies mentioned above arise from the wavelength coverage of their data-set, the definition of radiation field(s) they use or simply the dust model they choose. 

To reach coherent results on dust properties (dust-to-gas ratio, dust-to-metal ratio, fraction of PAHs, etc.), we need to be able to compare between model results.
In this paper, we carry out a rigorous comparison among some of the widely used dust models available in the literature by fitting the IR emission of M101 (NGC~5457) in a strictly identical way for all models. We therefore reduce the differences to those due to the physical modeling choices only. 
We compare the models from \citet[][]{Compiegne11}, \citet[][]{DL07}, and THEMIS \citep[overview in][]{THEMIS}, as well as \citet{Hensley+Draine_2020c}. We also include two modified blackbody models previously used in the literature: a simple-emissivity, and a broken-emissivity modified blackbody.

We perform our analysis on the galaxy M101 (NGC~5457), which has multiple advantages. 
First, its distance \citep[$\sim 6.7~$Mpc,][]{Tully09}, and low inclination allow for well-resolved photometry, even in the far-IR.
Second, the available IR photometry and spectroscopy from recent space telescopes with high sensitivity are ideal to constrain the dust models.
Finally, metallicity gradients of M101 also offer an independent route to put an upper limit on its dust content. The galaxy has detailed measurements of metallicity from auroral lines \citep[e.g.][]{Croxall16, Berg20}, which puts good constraints on the gas phase metal abundance. It has also been extensively targeted for deep \ion{H}{1} and \ion{CO}{0} observations \citep[][]{Walter08,Leroy09}, allowing us to account for all the gas (modulo any limitations in the ability of the CO and 21~cm lines to trace the gas). Combining these, we can evaluate the impact of model choice on derived dust properties across a wide range of environments with well-understood metal and gas content.

In Section~\ref{SecData} we present the photometry used to sample the IR emission of M101. Section~\ref{SecModeling} describes the physical and modified blackbody dust models and Section~\ref{SecMethodology} the technical aspects of the emission fitting technique. The results of these fits are presented in Section~\ref{SecResults} with discussions of the differences in dust properties yielded by the dust models. Finally in Section~\ref{SecAnalysis} we analyze the calibration differences in the dust models themselves.

\section{Data}
\label{SecData}
We present our adopted distance and orientation parameters for M101 in Table~\ref{TabM101}.

\renewcommand{\arraystretch}{1.1}
\begin{deluxetable}{lll}
\caption{M101 (NGC~5457) properties.}
\label{TabM101}
    \tablehead{
    Property & Value & Reference}
    \startdata
    \hline
    \hline
    Right ascension & 14:03:12.6 &  \citet[][]{Makarov14}\tablenotemark{a} \\
    Declination & +54:20:57 & \citet[][]{Makarov14}\tablenotemark{a} \\
    Inclination & 30\degree & \citet[][]{deBlok08}\tablenotemark{b}\\
    Position angle & 38\degree & \citet[][]{Sofue99} \\
    \rtf & $11.4'$ & \citet[][]{Makarov14}\tablenotemark{a} \\
    Distance & 6.7~Mpc & \citet[][]{Tully09}
    \enddata
\tablenotetext{a}{from the HyperLEDA data base; \url{http://leda.univ-lyon1.fr/}}
\tablenotetext{b}{Note the difference with the HyperLEDA database value (16\degree).}
\end{deluxetable}

In order to derive the dust properties in M101, we perform fits to its IR SED, comprised of measurements at 16 different photometric bands:
\begin{itemize}
\renewcommand\labelitemi{\tiny$\bullet$}
    \setlength\itemsep{0.1cm}
    \item 3.4, 4.6, 12, and 22~$\mu$m from the {\it Wide-field Infrared Survey Explorer} \citep[WISE;][]{Wright10}. We use the maps delivered by \citet[][]{Leroy19} in the $z=0$ Multiwavelength Galaxy Survey, already convolved to a $15''$-FWHM Gaussian resolution;
    \item 3.6, 4.5, 5.8, and 8.0~$\mu$m from the {\it Infrared  Array Camera} instrument \citep[IRAC;][]{Fazio04}, and 24 and 70~$\mu$m from the {\it Multiband Imaging Photometer for Spitzer} instrument \citep[MIPS;][]{Rieke04}\footnote{We do not use MIPS~160 \citep[e.g. as opposed to][]{Aniano20} to gain back some resolution (MIPS~160 PSF is $\sim 39''$), without losing wavelength coverage. This does not lead to major differences.}, both on-board the {\it Spitzer Space Telescope} \citep[][]{Werner04}. We used the product delivery DR5\footnote{\url{https://irsa.ipac.caltech.edu/data/SPITZER/LVL/LVL_DR5_v5.pdf}} from the Local Volume Legacy survey for IRAC and MIPS maps \citep[][]{Dale09};
    \item 70, 100, and 160~$\mu$m from the {\it Photoconductor  Array Camera and Spectrometer} instrument \citep[PACS;][]{Poglitsch10} and 250, 350, and 500~$\mu$m from the {\it Spectral and Photometric Imaging  Receiver} instrument \citep[SPIRE;][]{Griffin10}, both on-board the {\it Herschel Space Observatory} \citep[][]{Pilbratt10}. We downloaded the scans from the KINGFISH program \citep[][]{Kennicutt11} in the Herschel Science Archive, and processed them from L0 to L1 with \texttt{HIPE} \citep[v.~15; PACS Calibration~v.~77; SPIRE Calibration~v.~14.3;][]{Ott10} and \texttt{Scanamorphos} \citep[v. 25;][]{Roussel13}.
\end{itemize}{}

\subsection{Data Processing}
We correct the images for extended source calibration appropriately and convert them to the same units in the following way.
We apply extended-source correction factors for the IRAC images. These are multiplicative corrections of 0.91, 0.94, 0.695, and 0.74 at 3.6, 4.5, 5.8 and 8.0~$\mu$m, respectively, as suggested by the IRAC Instrument Handbook\footnote{\url{https://irsa.ipac.caltech.edu/data/SPITZER/docs/irac/iracinstrumenthandbook/29/}}.
For SPIRE, we adopt calibration factors of 90.646, 51.181, and 23.580 MJy/sr/(Jy/beam) at 250, 350, and 500~$\mu$m, respectively, as suggested by the SPIRE Handbook\footnote{\url{http://herschel.esac.esa.int/Docs/SPIRE/spire_handbook.pdf}}. These factors convert the data to MJy/sr from Jy/beam and also correct from point source to extended source calibration.
The PACS data in units of Jy/pixel are converted to units of MJy/sr using the image pixel size (contained in the headers).

We then process all the maps following the same steps, as described below.
We remove a background in each image at their native resolution, by fitting a 2-D plane using regions identified not to have significant galaxy emission. We find these regions with the following procedure: 1) We use the \rtf radius (\rtf $\sim 12'$) to first mask the galaxy (this covers the visible SPIRE~500 emission completely); 2) We measure the median of the remaining pixels and the standard deviation, and clip all of those that are 3 standard deviations above that median; 3) We iterate that clipping until the medians at iterations $i$ and $i+1$ differ by less than 1\%.
This clipping is only done for the purpose of measuring a background level, and we do not keep this mask applied to the data for the following steps.
Table~\ref{TabBkgs} lists the final standard deviations of the background pixels in each bands.

\renewcommand{\arraystretch}{1.1}
\begin{deluxetable}{lccc}
\caption{Band-related details. References to the $\sigma_{\rm cal}$ and $\sigma_{\rm sta}$ coefficients: WISE: \url{http://wise2.ipac.caltech.edu/docs/release/prelim/expsup/sec4_3g.html}; IRAC: \citet[][]{Reach05} and \url{https://irsa.ipac.caltech.edu/data/SPITZER/docs/irac/iracinstrumenthandbook/29/}; MIPS: \citet[][]{Engelbracht07MIPS24} and \citet[][]{Gordon07MIPS70}; PACS: \citet[][]{Muller11_PACS} and \citet[][]{Balog13}; SPIRE: \citet[][]{Griffin10} and \citet[][]{Bendo13}.}
\label{TabBkgs}
    \tablehead{
    Band & $\sigma_{\rm bkg}$\tablenotemark{a} & $m_{\rm cal}$\tablenotemark{b} & $m_{\rm sta}$\tablenotemark{c} \\
    & 10$^{-1}$ MJy/sr & \% & \% }
    \startdata
    \hline
    \hline
    WISE~3.4 & 0.170 & 2.4 & 10.0 \\
    IRAC~3.6 & 0.149 & 9.0 & 1.5 \\
    IRAC~4.5 & 0.107 & 6.0 & 1.5 \\
    WISE~4.6 & 0.095 & 2.8 & 10.0\\
    IRAC~5.8 & 0.109 & 30 & 1.5 \\
    IRAC~8.0 & 0.085 & 26 & 1.5 \\
    WISE~12 & 0.114 & 4.5 & 10.0 \\
    WISE~22 & 0.055 & 5.7 & 10.0 \\
    MIPS~24 & 0.049 & 4.0 & 0.4 \\
    MIPS~70 & 5.24 & 10.0 & 4.5 \\
    PACS~70 & 10.7 & 10.0 & 2.0 \\
    PACS~100 & 10.1 & 10.0 & 2.0 \\
    PACS~160 & 7.95 & 10.0 & 2.0 \\
    SPIRE~250 & 3.96 & 8.0 & 1.5 \\
    SPIRE~350 & 2.73 & 8.0 & 1.5 \\
    SPIRE~500 & 1.82 & 8.0 & 1.5 
    \enddata
\tablenotetext{a}{The background standard deviation in the pixels considered to measure the background covariance matrix, once the maps are background-removed, convolved and projected.}
\tablenotetext{b}{$m_{\rm cal}$ is the error on the calibration used in each instrument. The large errors of the IRAC bands are from the extended source corrections, which we consider to be correlated calibration errors.}
\tablenotetext{c}{$m_{\rm sta}$ measures the stability of an instrument, i.e. the scatter when measuring the same signal.}
\end{deluxetable}{}

\begin{figure*}
    \centering
    \includegraphics[width=\textwidth]{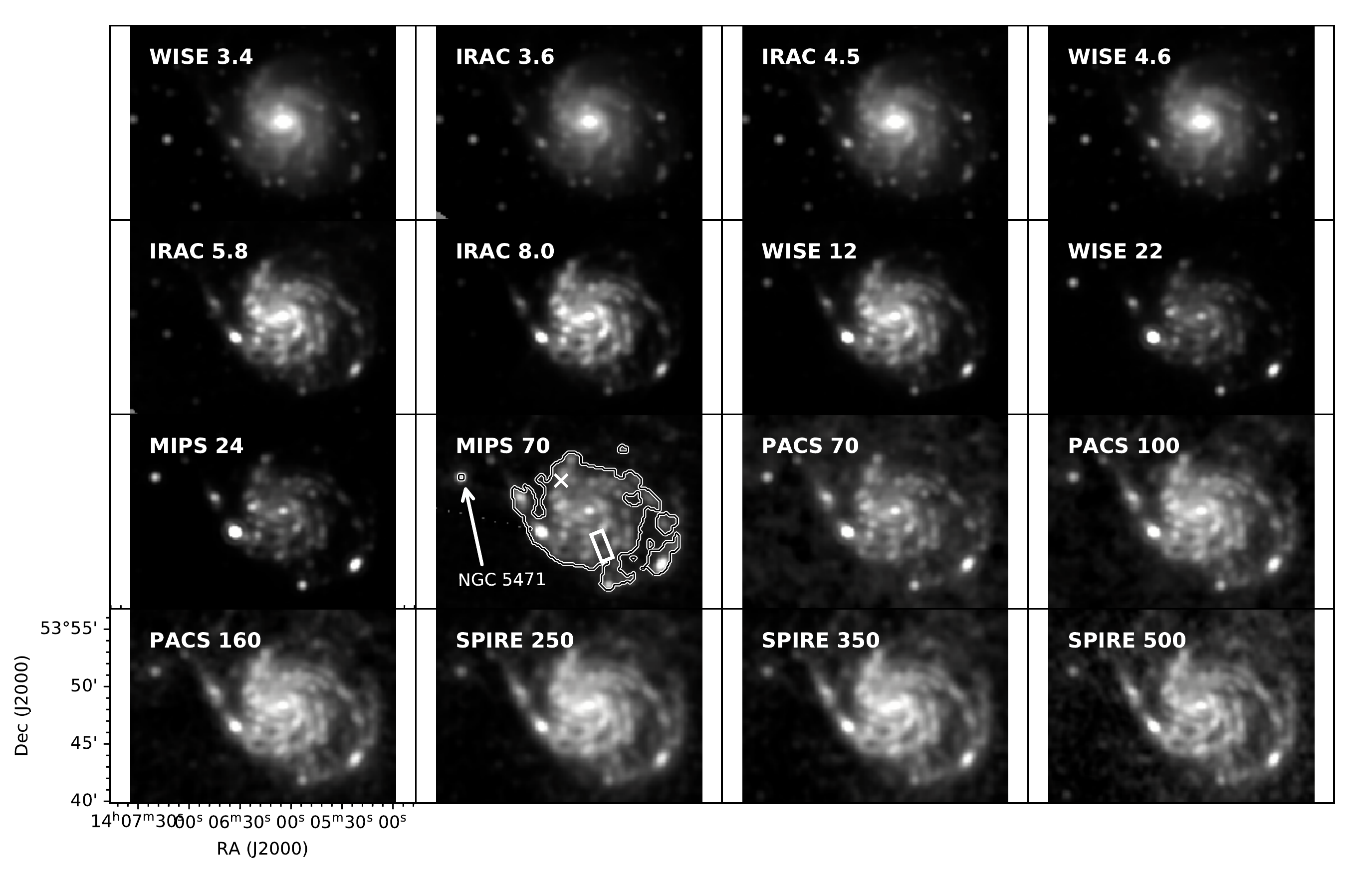}
    \caption{Emission of M101 (NGC~5457) from 3.4 to 500~$\mu$m, in all the bands used in this study. The maps show the final version of each map: after extended source correction (IRAC and SPIRE bands), unit conversion (PACS and SPIRE bands), background removal, convolution to SPIRE~500 PSF ($\sim 36''$) and regridding. On the MIPS~70 panel, we also show: the location of the pixel for the fit example (white cross; Figure~\ref{FigFitsPanel}), the region used for the IRS measurement (white rectangle; Section~\ref{SecMidIR}), and the contours for the $3\sigma$ detection threshold.} 
    \label{FigData}
\end{figure*}

Each map is then convolved to the SPIRE~500 PSF (FWHM~$\sim 36''$), using convolution kernels from \citet{Aniano11}.
Finally, all the maps are aligned and projected onto the astrometric grid of the SPIRE~500 image.
In Figure~\ref{FigData}, we show the 16 bands that we use to model the dust emission from 3.4 to 500~$\mu$m.

The final pixel size ($9''$) oversamples the SPIRE~500 beam size, to which all data are convolved. We take this into account by correcting by $\sqrt{N_{\rm pix}/N_{\rm beam}}$ when calculating uncertainties on quantities that use the values of multiple pixels.

{}

\subsection{Ancillary data: neutral and molecular gas}
In Section~\ref{SecAnalysis}, we use gas surface density and metallicity measurements in additional to IR SEDs. We use the same data as \citet[][]{Chiang18} to get the total gas surface density, $\Sigma_{\rm gas}$, combining gas surface density maps from \hi and \coto emission converted to H$_2$. 
The \hi line data is from the THINGS collaboration \citep[The \ion{H}{1} Nearby Galaxy Survey;][]{Walter08}. The map is converted from integrated intensity to surface density assuming optically thin 21~cm emission, following \citet[][Equ. 1 and 5, and multiplying by the atomic mass of hydrogen]{Walter08}.

We use ${\rm CO}~(2\to1)$ emission from HERACLES \citep[][]{Leroy09}, and convert it to H$_2$ surface density, assuming a line ratio $R_{21}=0.7$, and a $\alpha_{\rm CO}$ conversion factor, in units of M$_\odot$/pc$^2$ (K km/s$)^{-1}$,  following two prescriptions (both including helium): the first one is representative of the MW CO-to-H$_2$ conversion\footnote{A standard $X_{\rm CO} = 2\times10^{20}$~(K~km/s)$^{-1}$ is assumed for the column density conversion factor. The mass of helium and heavier elements have been accounted for in \aco.}, 
\begin{equation}
    \alpha_{\rm CO}^{\rm MW} = 4.35,
\label{EquMW}
\end{equation}
and the second one follows \citet*[][]{Bolatto13},
\begin{equation}
    \begin{split}
    \alpha_{\rm CO}^{\rm BWL13} = 2.9\ &\text{e}^{(0.4/Z')}(\Sigma^{100}_{\rm Total})^{-\gamma} \\
    &\textrm{with}
        \begin{dcases}
        \gamma = 0.5\ &\text{if } \Sigma^{100}_{\rm Total} \geq 1 \\
        \gamma = 0\ &\text{otherwise}
        \end{dcases}
    \end{split}
\label{EquBWL}
\end{equation}
where $Z'$ is the metallicity relative to solar metallicity\footnote{$Z' = 10^{\left ( \rm (12+log(O/H)) - (12+log(O/H)_\odot) \right ) }$ with 12+log(O/H)$_\odot=8.69$.}, $\Sigma^{100}_{\rm Total}$ is the total surface density map in units of 100~M$_\odot$/pc$^2$.
The gas maps are convolved to a $41''$ Gaussian PSF \citep[][]{Chiang20}; since we plot the radial profile of the gas maps by averaging in growing annuli, we find that the minor differences between the dust and gas maps resolutions are negligible.
We measure an uncertainty of $\sim~0.3$~K~km/s for the \coto map, and an uncertainty of $\sim~0.4~$\msolpc for the \hi map\footnote{The rms per channel is $\sim 0.46~$mJy/beam. Assuming $\sigma_{\rm z,~gas} = 11~$km/s \citep[][]{Leroy08}, the uncertainty is 0.96~\msolpc. The 5\% calibration error \citep[][]{Walter08} becomes significant in the dense regions.}. We build a radial profile by averaging $\Sigma_{\rm gas}$ in growing annuli from the center out to \rtf.

\subsection{Ancillary data: metallicity}
\label{SecDatMetal}
Metallicity measurements $12+{\rm log_{10}(O/H)}$ are taken from the CHAOS survey \citep[][]{Berg20}, derived from 72 \ion{H}{2} regions of M101. In particular, we use their fitted metallicity gradient
\begin{equation}
    12+{\rm log_{10}(O/H)} = 8.78 \pm 0.04 - 0.75 \pm 0.07\ (r/{\rm r_{25}})
\end{equation}
to convert it to a radial profile of metal surface density (see Section~\ref{SecMaxDust}). 
We adjusted the slope given by \citet[][0.90]{Berg20} to account for the different \rtf used here. 
\citet{Berg20} find that the scatter in individual region metallicities around the measured gradient is $\sim 10$\%.

Tracing the total metal mass from nebular emission lines of oxygen is subject to systematic uncertainties. As pointed at by \citet[][]{Jenkins09}, oxygen shows unexplained depletion patterns, compared to how much is expected to be locked in the solid phase.
However, \citet[][]{Peimbert10} measured the depletions of heavy elements in Galactic and extra-galactic \ion{H}{2} regions, and found minimal depletion of oxygen ($\lesssim 0.1~$dex).
Other elements may be of use for tracing metallicity \citep[as it is the case in][]{Berg20}, but given the widespread use of O/H for extragalactic studies, we use the typical $12+{\rm log(O/H})$ tracer.

\section{Dust emission models}
\label{SecModeling}
The goal of this study is to investigate the differences in dust properties derived using physical and modified blackbody models. In this Section, we present the characteristics of each model. The parameters we use to fit the IR SEDs are presented in Table~\ref{TabFitParams}. A summary table of the following information is presented in Appendix~\ref{AppModelTable}.

\subsection{Modified blackbodies}
\label{SecMBBMods}
We use single-temperature modified blackbodies models. For these two models, we only fit photometry from 100 to 500~$\mu$m. At $\lambda < 100~\mu$m, stochastically heated grains contribute to the emission and are not well modeled by a modified blackbody. Assuming the optically thin case, the dust emission $I_\nu$ is described as
\begin{equation}
I_\nu(\lambda) = \kappa_\nu(\lambda)\ \Sigma_{\rm dust}\ B_\nu(\lambda, T_{\rm dust}),    
\end{equation}
where $B_\nu(\lambda, T_{\rm dust})$ is the Planck function at wavelength $\lambda$ (in MJy/sr), \td the dust temperature, \sigd the dust surface density, and $\kappa_\nu$ the opacity. 
We use the simple power-law opacity (Equ.~\ref{EquKappaSE}) and a broken power-law opacity (Equ.~\ref{EquKappaBE}), which we normalize at $\lambda = 160~\mu$m. The broken power-law model presented the best results in terms of quality of fits in the study by \citet{Chiang18} and yielded physically reasonable \sigd and \td values in their study. 

We follow \citet[][]{Gordon14} and \citet{Chiang18} and calibrate the opacity values for each model.
We fit the modified blackbody model to the dust emission per H column of the MW high-latitude cirrus  described in \citet[][]{Chiang18} to derive the opacity. 
The abundance constraints are based on a depletion strength factor typical for lines-of-sight with similar $N_{\rm H}$ to the MW cirrus, e.g., $F_*=0.36$ \citep[][]{Jenkins09}. This sets the allowed dust mass per H atom. 
By fitting the temperature for the MW cirrus we can then derive the opacity for each modified blackbody model.
More details on the opacity and comparison to the physical models can be found in Section~\ref{SecOpacities}.

\subsubsection{Simple-Emissivity (SE)}
In this case, the opacity is described as a single power-law:
\begin{equation}
\kappa_\nu = \kappa_{\nu_0} \left (\frac{\nu}{\nu_0} \right )^\beta,
\label{EquKappaSE}
\end{equation}{}
\noindent where $\beta$ is the spectral index.
For all blackbody models, we fix $\lambda_0 = 160~\mu$m, and $\nu_0 = c/\lambda_0$.
The free parameters for this model are then the dust surface density, \sigd, the dust temperature, \td, and the dust spectral index, $\beta$. In this model the calibrated $\kappa_{\nu_0}=10.10 \pm 1.42$~cm$^2$/g \citep{Chiang18}, from fitting the high latitude cirrus as described above.

\subsubsection{Broken-Emissivity (BE)}
The \BE model stemmed from the identification of the sub-millimeter excess \citep[][]{Gordon14}. It allows a change of the dust spectral index with wavelength, meant to better reproduce the far-IR slope than does a simple modified blackbody.
In this model, the value of the opacity is wavelength-dependent, such that:
\begin{equation}
\kappa_\nu=\left\{\begin{array}{ll}
    \kappa_{\nu_0} \left ( \frac{\nu}{\nu_0} \right )^{\beta} & \lambda < \lambda_c \\
    \kappa_{\nu_0} \left ( \frac{\nu_c}{\nu_0} \right )^{\beta} \left ( \frac{\nu}{\nu_c} \right )^{\beta_2} & \lambda \geq \lambda_{\rm c}
    \end{array}\right.
\label{EquKappaBE}
\end{equation}
where \lambdac is the wavelength at which the opacity changes, and $\nu_{\rm c}$ its equivalent frequency. Following \citet[][]{Chiang18}, we fix $\beta=2$, and $\lambda_{\rm c} = 300~\mu$m. The free parameters are then the dust surface density, \sigd, the dust temperature, \td, and the second dust spectral index, $\beta_2$. The calibrated opacity, $\kappa_{\nu_0}$, for this model is $20.73 \pm 0.97$~cm$^2$/g \citep{Chiang18}, as described above.

\subsection{Opacity calibrations}
\label{SecOpacities}
The physical dust models used in our study have opacities that are set by each individual model’s calibration procedure. All models use similar constraints from the MW high latitude cirrus (described in more detail in Appendix A). For ease of comparison to the opacities we have derived for the modified blackbody models (10.1~cm$^2$/g for the \SE and 20.7~cm$^2$/g for the \BE at 160~$\mu$m), we list the opacity in the physical models below all scaled to 160 microns for comparison:
\begin{itemize}
\setlength\itemsep{0.02cm}
    \item \citet[][]{DL07}: from the \citet[][]{WD01Ext} model updated in \citetalias[][]{DL07} 
    we report $\kappa_{160}~=~10.2~$cm$^2$/g \citep[see also][]{Bianchi13};
    \item \citet[][]{Compiegne11}: from the work of \citet[][]{Bianchi13}
    we report $\kappa_{160}~=~12.0~$cm$^2$/g;
    \item \citet[][THEMIS]{THEMIS}: from the work of \citet[][]{GallianoReview18} 
    we report $\kappa_{160}~=~14.2~$cm$^2$/g;
    \item \citet[][]{Hensley+Draine_2020c}: the models uses $\kappa_{160}~\sim~9.95~$cm$^2$/g (B. Hensley; priv. comm.).
\end{itemize}

For all models, we use the listed opacity regardless of the specific environment in M101 we are studying.
Few of the currently available physical dust models have been calibrated in any other environment than the MW cirrus \citep[although the \citeauthor{DL07} \citeyear{DL07} model does have Small and Large Magellanic Cloud-like calibrations, though they are not widely used even for these very galaxies;][]{Sandstrom10, Chastenet19}.
Indeed, it is standard in current extragalactic applications to apply MW cirrus $R_V=3.1$ dust calibrations across all environments \citep[e.g.][]{Davies17, Aniano20}. In detail, this is unlikely to be correct since the opacity can and probably should evolve as a function of environment and it is clear that a single $F_*~=~0.36$ value does not describe the depletion in the ISM over the full range of column densities probed in galaxies. 
However, for the purposes of our comparative study of widely used dust models, we proceed by using the $R_V=3.1$ MW cirrus calibrations from each model.
Even if there were a potential way to adjust opacity with H column in M101, work by \citet[][]{Ysard15} using \citetalias[][]{THEMIS} suggest that this is insufficient to predict changes in dust properties.

\subsection{Physical dust models}
\label{SecPhysMods}
Physical dust models assume a composition, density, and shape for the dust grains and adopt heat capacities and optical properties from laboratory and theoretical studies that are appropriate for such materials. For simplicity, most models assume spherical grains or planar molecules for PAHs. In the case of PAHs, the grains/molecules are additionally described by an ionization state, which changes the absorption cross-sections as a function of wavelength. The temperature and emission of a grain of a given size, shape and composition in a radiation field with a specified intensity and spectrum can then be calculated analytically \citep[e.g.][]{DraineLee84,Desert1986}. 

The full dust population is represented by a grain size distribution and abundance relative to H for the specified compositions. Physical dust models are calibrated by adjusting the grain size distributions and dust mass per H to simultaneously match observations of extinction, emission, and abundances (and more recently, polarization) in a location where the underlying radiation field that is heating the dust is well known. This has generally been taken to be the high-latitude MW cirrus, where the radiation field intensity and spectrum are approximately given by the \citet[][]{Mathis83} model for the Solar neighborhood. 
The degree to which the models must adhere to the somewhat uncertain abundance constraints varies model to model. 
For example, the modified blackbody models from \citet[][]{Gordon14} use the depletion measurements in the MW as a strict limit.
However, most physical models allow the final element abundances to vary from depletions, using them only as loose guide. 

In the following, we use 4 physical dust models: \citet[][]{DL07}, \citet[][]{Compiegne11}, THEMIS \citep[][]{THEMIS} and \citet[][]{Hensley+Draine_2020c}. Here we briefly describe these models, the key differences between them. Details on their respective calibration methodologies can be found in Appendix~\ref{AppModelTable}.

\subsubsection{\citet[][]{DL07}}
In the \citet[][hereafter DL07]{DL07} model, dust is comprised of PAHs, graphite grains and amorphous silicate grains. It stems from the original models presented in \citet[][]{LD01}. 
The carbonaceous dust optical properties are adopted from \citet[][]{LD01PAH} with updates to the PAH cross-sections and form of the grain size distribution. A balance between ionized and neutral PAHs is assumed, following \citet[][]{LD01PAH}. The optical properties of silicate material are adopted from the ``astrosilicates'' in the original model. We do not make use of the mass renormalization in \citet[][]{Draine14}.

The mass fraction of PAHs, \qpah, is described as the mass of carbonaceous grains with less than 10$^3$ carbon atoms with respect to total dust mass. We effectively obtain \qpah in \% by converting the part-per-million carbon abundance $b_{\rm C}$ to a PAH fraction, using the reference $q_{\rm PAH} = 4.7\% \equiv b_{\rm C}=55~{\rm ppm}$.

The calibration of the model is described in several papers \citep[][]{DL01SamC, LD01PAH, WD01Ext}. We use the \citetalias{DL07} Milky Way model, with $R_{\rm V}=3.1$, which has a fixed ratio of silicate to carbonaceous grains.
Details on the calibration can be found in Appendix~\ref{AppDL07}.

\subsubsection{\citet[][]{Compiegne11}}
\label{SecModelMC11}
The \citet[][hereafter MC11]{Compiegne11} model is composed of PAHs, hydrogenated amorphous carbon grains, and amorphous silicate grains. The size distribution of the carbonaceous components includes PAHs, small amorphous carbon grains (SamC), and large amorphous carbon grains (LamC).
The PAH cross-sections and ionization as a function of size adopted in the model are based on \citet[][]{DL07} with slight modifications to the cross sections of several bands.  Amorphous carbonaceous grains have optical properties from \citet[][]{Zubko04} and heat capacities from \citet[][]{DL01SamC}. The amorphous silicates (aSil) have optical properties from \citet[][]{Draine03} and heat capacities from \citet[][]{DL01SamC}. 
Details on the calibration can be found in Appendix~\ref{AppMC11}.

The \dustem\footnote{\dustem is a tool that outputs extinction, emission and polarization of dust grains heated by a given radiation field. See details at \url{https://www.ias.u-psud.fr/DUSTEM/}.} tool allows both ionized and neutral PAHs to be fit independently.
Because not all models allow that separation, we tie their emission spectra together by summing them, hence keeping their ratio constant at roughly 60\% neutral and 40\% ionized\footnote{The ratio of ionized and neutral PAHs is size-dependent. The values are set in \dustem from the \texttt{MIX\_PAH\textit{x}.DAT} files.}.
Additionally, we fix the mass fractions of the large carbonaceous-to-silicate grains such as $(M_{\rm dust}^{\rm LamC}/M_{\rm H}) / (M_{\rm dust}^{\rm aSil}/M_{\rm H}) = 0.22$\footnote{This value is that from the \dustem file \texttt{GRAIN\_MC10.DAT}.}. Since these share a very similar far-IR spectral index, their respective emission cannot be properly determined independently with our wavelength coverage, and would lead to degenerate abundances if fit separately.

\subsubsection{\citetalias[][]{THEMIS}}
The Heterogeneous dust Evolution Model for Interstellar Solids \citep[THEMIS;][]{Jones13, Kohler14, Ysard15, THEMIS} is a core/mantle dust model, consisting of large silicate and hydrocarbons, both coated with aromatic-rich particles (Hydrogenated Amorphous Hydrocarbons, HACs). This model defines its dust components focusing strongly on laboratory data, slightly adjusted to better match observations. We use the diffuse ISM version of the model, described in \citet[][]{Jones13}.
The amorphous carbon grain properties are size-dependent, and there is no strictly independent population of grains responsible for the mid-IR features, carried by aromatic clusters in the form of mantles and very small grains (sCM20). As they grow to larger grains (lCM20) in size, their core becomes aliphatic-rich, coated with an aromatic mantle.
The silicate grains (aSilM5) in particular are discussed in \citet[][]{Kohler14}. They are a mixture of olivine and pyroxene-type material, with nano-inclusions of Fe and FeS. Their mass ratio is kept constant due to their extreme resemblance in emission in the considered wavelength range. 
The dust evolution models (from diffuse to denser medium) are discussed in \citet[][]{Ysard15}, with the impact of aggregates and thicker mantles on the final abundances. In the diffuse model we use, aSil have a 5~nm mantle, and hydrocarbons have a 20~nm mantle.
In our fitting, the lCM20-to-aSilM5 mass ratio is kept constant, such as $(M_{\rm dust}^{\rm lCM20}/M_{\rm H}) / (M_{\rm dust}^{\rm aSilM5}/M_{\rm H}) = 0.24$\footnote{This value is that from the \dustem file \texttt{GRAIN\_J13.DAT}}.
Details on the calibration can be found in Appendix~\ref{AppTHEMIS}. 

\subsubsection{\citet[][]{Hensley+Draine_2020c}}
Rather than employ separate amorphous silicate and carbonaceous grain components, the \citet[][hereafter HD21]{Hensley+Draine_2020c} model invokes a single homogeneous composition, ``astrodust'' \citep{Draine+Hensley_2020a}, to model most of the interstellar grain mass. In addition to astrodust, the model incorporates PAHs using the cross-sections from \citet{DL07} and a small amount of graphite using the turbostratic graphite model presented in \citet{Draine_2016}. The \citetalias{Hensley+Draine_2020c} model was developed to reproduce the observed properties of dust polarization, including both polarized extinction and emission, in addition to total extinction and emission. This results in raising of the emissivity in the far-IR, forcing the dust to be slightly cooler than other models and requiring a higher radiation field to get comparable amounts of emission.
We use cross sections computed for 2:1 prolate spheroids, but the grain shape has only a small effect on the far-infrared total intensity studied in this work.
Details on the calibration can be found in Appendix~\ref{AppHD20}.

For the purposes of this study, the \citetalias{Hensley+Draine_2020c} model has been parameterized in the same way as \citet[][]{DL07}, i.e., utilizing parameters $U$ and $q_{\rm PAH}$. 

\section{Fitting methodology}
\label{SecMethodology}
\subsection{Making matched model grids}
Since each model was developed independently, it is not always possible to create model grids that have exactly the same parameter sampling, because of the lack of parameter equivalence. However we attempt to do so as much as possible.

\textit{Radiation field --- }
We implement identical dust heating in each of the physical models: a fraction $\gamma$ of the dust mass in a pixel is heated by a power-law distribution of radiation fields with $U_{\rm min} < U \leq {\rm U_{max}}$ \citep{Dale01}, where $U$ is the interstellar radiation field, expressed in units of the MW-diffuse radiation field at 10~kpc from \citet[][]{Mathis83}. The remaining fraction of dust $(1-\gamma)$ is heated by the minimum radiation field \umin \citep{DL07}:
\begin{equation}
\begin{split}
    \frac{1}{M_{\rm dust}} \frac{dM_{\rm dust}}{dU} = 
    &(1-\gamma)\ U_{\rm min} + \\
    &\gamma\ \frac{(\alpha-1)}{U_{\rm min}^{1-\alpha}-{\rm U}_{\rm max}^{1-\alpha}}\ U^{-\alpha}
\end{split}{}
\label{EquRF}
\end{equation}{}
\noindent We fix $\alpha=2$ for all models, ${\rm U_{max}}=10^7$ for the \citetalias{DL07}, \citetalias{THEMIS} and \citetalias{Compiegne11} models, and ${\rm U_{max}}=10^6$ for \citetalias[][]{Hensley+Draine_2020c}. 
This choice is constrained by the available parameter ranges in the models: the ${\rm U_{max}}$ values for the \citetalias[][]{DL07} and \citetalias[][]{Hensley+Draine_2020c} models are fixed, and we do not have the freedom to change them; thanks to \dustem, we can adjust ${\rm U_{max}}$ for \citetalias[][]{Compiegne11} and \citetalias[][]{THEMIS}\footnote{Using \dustem and \citetalias[][]{THEMIS}, we checked the effect of using ${\rm U_{max}}=10^6$ or $10^7$, for a range of $\gamma$ values. We find that most of the mid-IR bands used in this study are only minimally affected. The difference for IRAC~4.5 and WISE~4.6 can be significant at $\gamma \geq 0.1$, while the maximum $\gamma$ values reach by the model fits is 0.07 (for \citetalias[][]{DL07}).
Additionally, these bands are dominated by starlight, and mostly modeled by the 5,000~K blackbody.}.
For \citetalias{THEMIS} and the \citetalias{Compiegne11} model, which have multiple dust populations that can be independently heated, each population is heated by the same radiation field.

The minimum radiation field parameter grid is fixed by the values provided in the \citet[][]{DL07} model.
Thanks to the \dustem\ tool, we can use the exact same values with \citetalias{THEMIS} and the \citetalias{Compiegne11} model. 
These parameter values in \citetalias[][]{Hensley+Draine_2020c} however are not exactly identical to those in $U_{\rm min}^{\rm DL07}$. We therefore use the spectra with closest $U_{\rm min}^{\rm HD21}$ values. 
Although not strictly equal, the radiation field values in Hensley \& Draine are within 5\% of $U_{\rm min}^{\rm DL07}$.

For the modified blackbody models, we use a single radiation field intensity and translate it into a dust temperature. We use the relationship 
\begin{equation}
U \propto T_{\rm dust}^{\beta + 4} \quad \textrm{ with } \beta = 2
\label{EquTpropU}
\end{equation}
to convert \umin to \td and find the approximately matching sampling to use in the blackbody models. We use the normalization from \citet[][]{Draine14}, i.e. $U=1$ at $T_{\rm dust} = 18~$K, found using the same radiation field from \citet[][]{Mathis83}.

\textit{Mid-IR feature carriers --- }
The \qpah parameter is kept strictly identical between \citetalias{DL07} and \citetalias{Hensley+Draine_2020c}.
We choose to use \citetalias{THEMIS} and the \citetalias{Compiegne11} models in a similar fashion. Despite the definition of ``PAHs'' being different in \citetalias{THEMIS}, we parameterize the model so that the fraction of small grains can vary, keeping the large-carbonaceous-to-silicate grains ratio constant.
The \citetalias[][]{Compiegne11} default model has 4 populations that can vary independently, and we choose to tie together SamC, LamC and aSil, to leave the amount of PAHs as a free parameter. 
We explain these choices in more detail in Section~\ref{SecDiscussion}.

\textit{Stellar surface brightness --- }
In addition to the dust model parameters, we use a scaling parameter of a 5,000~K blackbody, $\Omega_*$, to model the stellar surface brightness visible at the shortest wavelengths. This temperature is a good approximation as the shortest wavelengths are nearly on the Rayleigh-Jeans tail. The free parameter $\Omega_*$ scales the amplitude of the stellar blackbody.

\renewcommand{\arraystretch}{1.1}
\begin{deluxetable}{llll}
\caption{Fitting parameters}
    \centering
    \tablehead{
    \colhead{Parameter} & \colhead{Range} & \colhead{Step} & \colhead{Unit}} 
    \startdata
    \hline
    \hline
    \multicolumn{4}{c}{All physical models} \\
    \hline
    \logt(\sigd) & [-2.2, 0.1] & 0.035 & \msolpc \\
    \umin & [0.1, 50] & Irr\tablenotemark{a} & -- \\
    \logt($\gamma$) & [-4, 0] & 0.15 & -- \\
    \logt(\ostar) & [-1, 2.5] & 0.075 & -- \\
    \hline
    \multicolumn{4}{c}{Model-specific} \\
    \hline
    DL07, HD21: \qpah & [0, 6.1] & 0.25 & \% \\
    THEMIS: $f_{\rm sCM20}$\tablenotemark{b} & [0, 0.5] & 0.03 & -- \\
    MC11: $f_{\rm PAHs}$\tablenotemark{b} & [0, 0.5] & 0.03 & -- \\
    \hline
    \hline
    \multicolumn{4}{c}{Modified-blackbody models} \\
    \hline
    \logt(\sigd) & [-2.2, 0.1] & 0.035 & \msolpc \\
    \td & [12, 35] & 0.3 & K \\
    SE: $\beta$, BE: $\beta_2$ & [-1,4] & 0.05 & --
    \enddata
\tablenotetext{a}{$U_{\rm min} \in $ \{0.1, 0.12, 0.15, 0.17, 0.2, 0.25, 0.3, 0.35, 0.4, 0.5, 0.6, 0.7, 0.8, 1.0, 1.2, 1.5, 1.7, 2.0, 2.5, 3.0, 3.5, 4.0, 5.0, 6.0, 7.0, 8.0, 10.0, 12.0, 15.0, 17.0, 20.0, 25.0, 30.0, 35.0, 40.0, 50.0\}.}
\tablenotetext{b}{We use a ``fraction'' parameter so that $\Sigma_{X} = f_X \times \Sigma_{\rm d}$, where $X$~=~\{${\rm sCM20^{THEMIS}}$, ${\rm PAHs^{MC11}}\}$.}
\label{TabFitParams}
\end{deluxetable}{}

\begin{figure*}
    \centering
    \includegraphics[width=\textwidth]{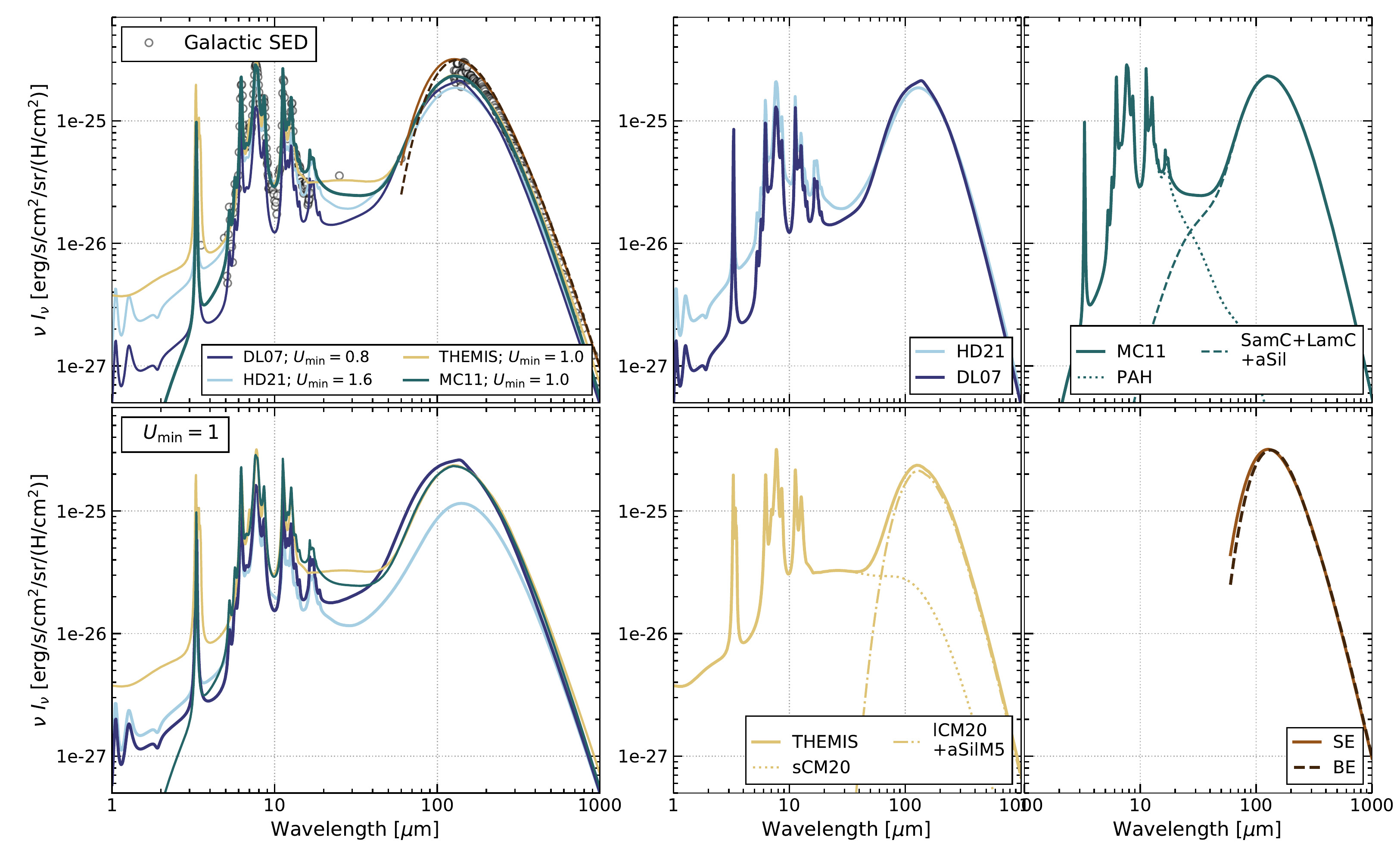}
    \caption{The dust emission models used in this study: \citet{DL07}, \citet[][THEMIS]{THEMIS}, \citet[][]{Compiegne11}, \citet{Hensley+Draine_2020c}, \SE, and \BE. {\it Top left:} all six models at the \umin value that best fits the calibration SED. {\it Bottom left:} the four physical models at \umin~=~1. The right side panels show each model at the \umin value that best fits the calibration SED, and their break-down as they are used in this study. }
    \label{FigModels}
\end{figure*}

In Table~\ref{TabFitParams}, we list the final free parameters we use for each model. There are 5 free parameters for the physical models, and 3 for the modified blackbody models.
Figure~\ref{FigModels} shows all the dust emission models used in this study. The top-left panel shows the fiducial MW high galactic latitude diffuse ISM models, labeled `Galactic SED'. The IR MW diffuse emission from \citet[][]{Compiegne11} is also plotted. The bottom-left panel shows the physical dust models at the same radiation field \umin~$=1$.
The other different panels detail the models: \citetalias{THEMIS} is divided in two grain population when fitted to the SED of M101 (note that we tie the lCM20 and aSilM5 population); for the \citetalias{Compiegne11} model we tie the SamC, LamC and aSil emissions together, and thus there are two free parameters for the dust mass: \fpah or $f_{\rm sCM20}$, and the total dust surface density.
The two modified blackbodies, are shown with their best fit values to the MW diffuse SED: \{\td~$=20.9$~K, $\beta=1.44$\} for the \SE model, and \{\td~$=18.0$~K, $\beta_2=1.55$\} for the \BE model.

{}

\subsection{Fitting tool}
\subsubsection{Bayesian fitting with \dustbff}
We use the \dustbff tool from \citet[][]{Gordon14} to fit the data with the chosen dust models. \dustbff is a Bayesian fitting tool that uses flat priors for all parameters. The probability that a model with parameters $\theta$ fits the measurements ($S^{\rm obs}$) is given by:
\begin{equation}
    \mathcal{P}({\bf S}^{\rm obs}|\theta) = \frac{1}{Q}\text{e}^{-\chi^2(\theta)/2}
\end{equation}
with
\begin{equation}
\begin{split}
    Q^2 &= (2\pi)^{\rm n}\ \text{det}|\mathbb{C}| \\
    \chi^2(\theta) &= [{\bf S}^{\rm obs} - {\bf S}^{\rm mod}(\theta)]^{\rm T}\ \mathbb{C}^{-1}\ [{\bf S}^{\rm obs} - {\bf S}^{\rm mod}(\theta)]
\end{split}
\end{equation}
where ${\bf S}^{X}$ the observed ($X={\rm obs}$) or modeled ($X={\rm mod}$) 16 band SEDs used here, and $\mathbb{C}$ is the covariance matrix that includes uncertainties from random noise, astronomical backgrounds and instrument calibration (described further below).

To create ${\bf S}^{\rm mod}(\theta)$ each model spectrum is convolved with the spectral responses for the photometric bands used here. PACS and SPIRE band integrations are done in energy units, whereas all the others are done in photon units, as necessitated by the instrument's calibration scheme.
We also follow the reference spectra used for the calibration of each instrument: the MIPS bands require a reference shape in the form of the $10^4$~K blackbody while the other bands use a reference shape as $1/\nu$.

\subsubsection{Covariance matrices}
\label{SecBkg}
The covariance matrix in the previous equations describes the uncertainties on the measured flux, both due to astronomical backgrounds, and instrumental noise and the uncertainties in calibrating the instruments.
This takes into account the correlation between the photometric bands due to the calibration scheme and the correlated nature of astronomical background signals. We define a background matrix and an instrument matrix such that $\mathbb{C} = \mathbb{C}_{\rm bkg} + \mathbb{C}_{\rm ins}$ to propagate the correlated errors of the background and noise, and of the calibration uncertainties, respectively.

\paragraph{Background covariance matrix $\mathbb{C}_{\rm bkg}$}
The ``background'' in our images encompasses astronomical signals that do not come from the IR emission of the target M101. It is dominated by different objects depending on the wavelength, and can therefore be correlated between bands. For instance, the background from 3.4 to 5.8~$\mu$m is mostly from foreground stars, as well as zodiacal light; from 8 to 24~$\mu$m, it is from evolved stars and background galaxies; in the far-IR, it is dominated by Galactic cirrus emission and background galaxies.
To include this uncertainty, we measure this combination of signals in ``background pixels'' using the processed data and a masking procedure described in Section~\ref{SecStarsBkg}.

The elements of the background covariance matrix are calculated as
\begin{equation}
    (\mathbb{C}_{\rm bkg})_{i,j}^2 = \frac{\sum_k^{\rm N}{(S_i^k - \langle S_i \rangle)(S_j^k - \langle S_j \rangle)}}{{\rm N}-1},
\end{equation}
where $S_{X}^k$ is the flux of pixel $k$, in band $X$, and $\langle S_{X} \rangle$ is the average background emission in band $X$ (close to 0). The final number of 9$''$ pixels used to measure the covariance matrix is $N=18,920$.

\begin{figure}
    \centering
    \includegraphics[width=0.5\textwidth, trim={0 0 0 0.7cm}, clip]{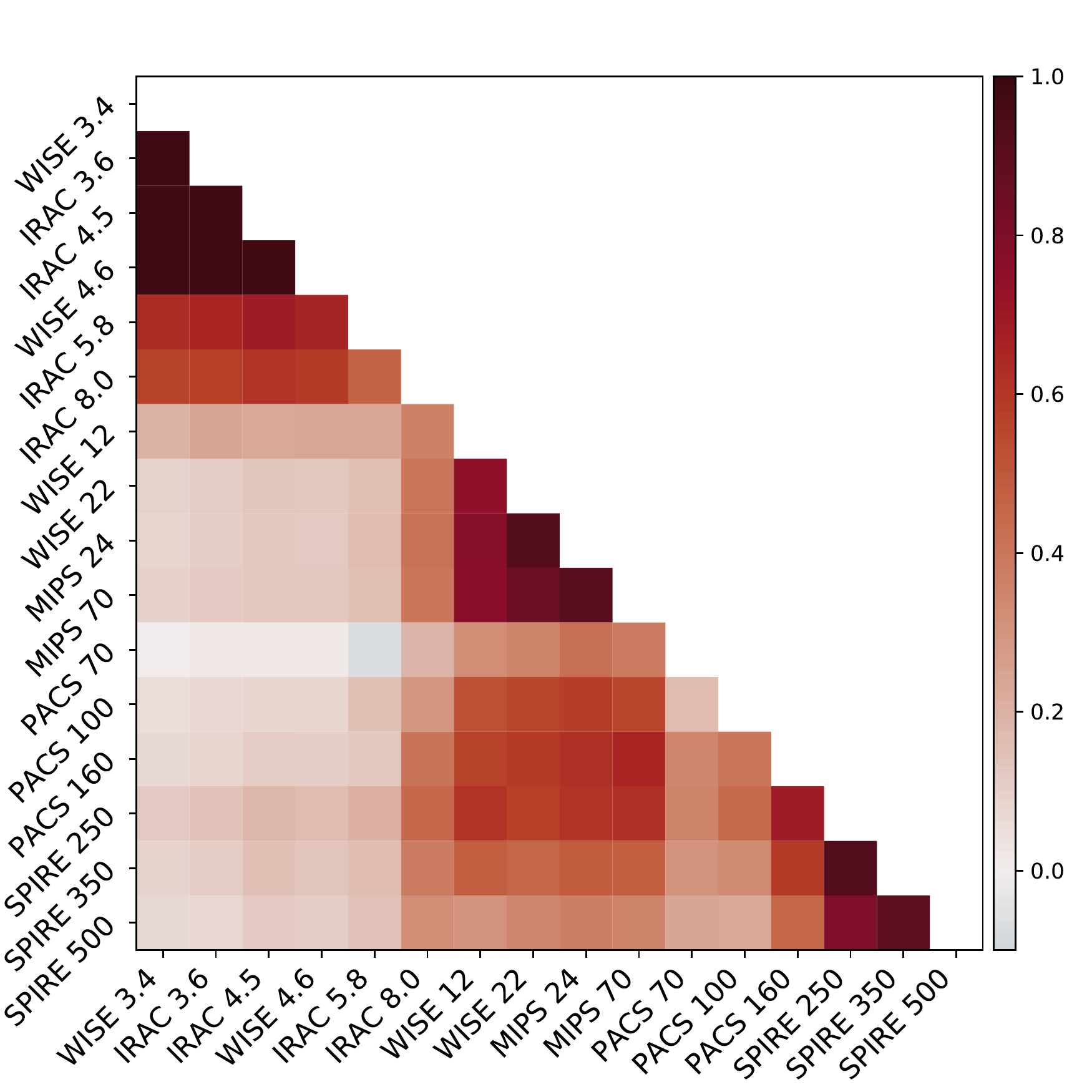}
    \caption{Background correlation matrix. The dark color indicates a strong correlation between the bands. The matrix is symmetric and we show only half.
    The text indicates the astronomical signals that dominates the contoured bands, and that explains their strong correlation.}
    \label{FigBkgMat}
\end{figure}

We display the background \emph{correlation} matrix in Figure~\ref{FigBkgMat} \citep[not the elements' absolute values but the Pearson correlation coefficients; see][]{Gordon14}. 
Three clear sets of positive correlations appear, like previously explained: correlations are due to starlight in the near-IR, evolved stars and MW cirrus in the mid-IR, and background galaxies and MW cirrus in the far-IR.
Additionally, some bands may be noise dominated: it is the case for the PACS~70 band which is only weakly correlated with other bands.

{}

\paragraph{Instrument calibration matrix $\mathbb{C}_{\rm ins}$}
This matrix is calculated as the quadratic sum of the correlated and uncorrelated errors for each instrument. The correlated errors (matrix with diagonal \emph{and} anti-diagonal terms) refer to the instrument calibration itself ($m_{\rm cal}$ in Table~\ref{TabBkgs}, while the uncorrelated (matrix with diagonal terms only) express the instrument stability, or repeatability ($m_{\rm sta}$ in Table~\ref{TabBkgs}).
We use the same calibration errors reported in \citet[][see Table~\ref{TabBkgs}]{Chastenet17} for the \spitzer/MIPS and \textit{Herschel} bands. 
The correlated errors for the IRAC bands were changed to take into account the uncertainties in the extended source correction factors, larger than the calibration error themselves. The errors due to repeatability are unchanged.
The errors for WISE bands were taken from the WISE documentation\footnote{\url{http://wise2.ipac.caltech.edu/docs/release/prelim/expsup/sec4_3g.html}}.

The elements of the calibration matrix  $\mathbb{C}_{\rm ins}$ are calculated ``model-by-model'' as
\begin{equation}
    (m_{\rm ins})_{i,j}^2 = S_i^{\rm mod}(\theta)\ S_j^{\rm mod}(\theta)\ (m^2_{{\rm cal,\ }i,j} + m^2_{{\rm sta,\ }i,j})
\end{equation}
with particular elements
\begin{equation*}
\begin{split}
    m_{{\rm cal,\ }i,j} &= 0\ \text{if }i,~j\text{ belong to two \emph{different} instruments;} \\
    m_{{\rm sta,\ }i,j} &= 0\ \text{if }i \ne j.
\end{split}
\end{equation*}

We fit all the pixels that are above 1$\sigma$ of the background values, in all bands. We use these pixels to show parameter maps and radial profiles, while the galaxy-integrated values are calculated for pixels above 3$\sigma$ detection above the background (black contours in Appendices~\ref{AppPrms} and \ref{AppResd}). 

\subsubsection{Stars in the background/foreground}
\label{SecStarsBkg}
Here we describe the masking procedure to measure the background covariance matrix in Section~\ref{SecBkg}. To do so, we use the final images, i.e. background-subtracted, convolved and projected to the same pixel grid.

The covariance matrix elements are calculated with the assumption of a Gaussian noise. While the assumption works well for faint and unresolved stars and the cosmic infrared background galaxies, it is no longer correct if we include bright stars.
Bright foreground stars only must therefore be cut to measure this matrix.
This masking has the effect of making the approximation of Gaussian noise for the remaining background more correct.
Note, however, than they are not masked for the fitting within the boundary of galaxy\footnote{We find no pixels showing a bad fit due to a foreground star. The $\Omega_*$ maps do not show conspicuous peaks, indicating that the foreground stars are not dominant, and the models successfully fit the galaxy emission.}, but only for the purpose of the covariance matrix measurement.

In the process of masking, we first exclude the region within \rtf, to mask the galactic emission.
We then mask the brightest stars, using the star masks from \citet[][]{Leroy19} that leverage the known positions of stars from the GAIA and 2MASS catalogs\footnote{\url{https://irsa.ipac.caltech.edu/data/WISE/z0MGS/overview.html}}. To match the final products, we convolve and regrid these star masks to the SPIRE~500 resolution. After convolution, the mask values are no longer binary 0 and 1, but show intermediate values around the position of bright stars. We mask pixels above 0.15 to exclude these regions where bright foreground stars contaminate our measurements from the covariance matrix calculation.

One argument against the decision to mask bright stars to measure the covariance matrix would be that the emission from these stars is an astrophysical signal that should be taken into account to propagate the noise from a band to another. This would require creating a new noise model and significant changes to the fitting methodology.
Rather than major changes to the fitting approach, we decide to mask the stars to measure the background covariance matrix.

\section{Results}
\label{SecResults}
\begin{figure*}
    \centering
    \includegraphics[trim={2cm 1cm 3cm 2.2cm}, clip, width=\textwidth]{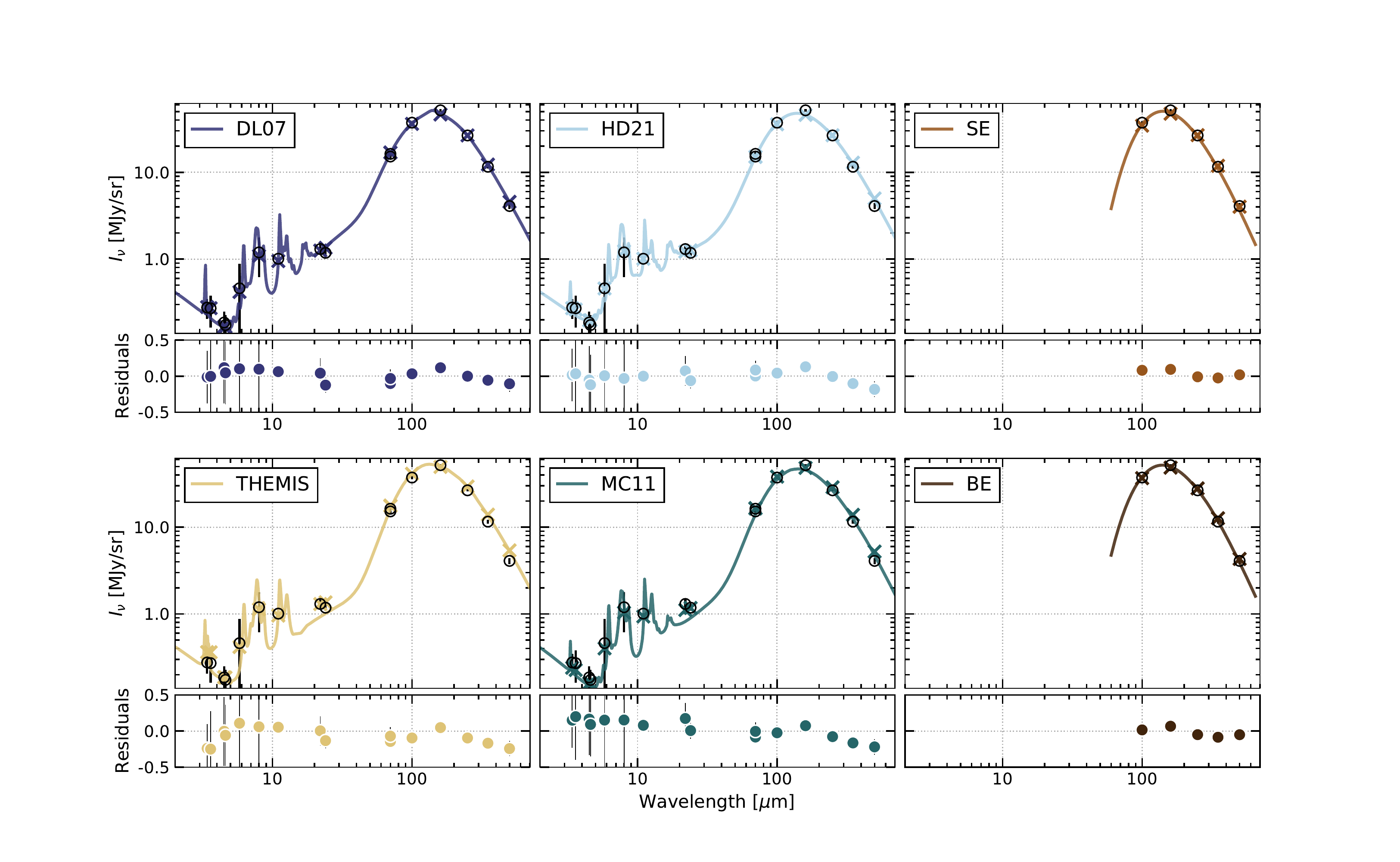}
    \caption{Example of the best fits in a pixel for the six models used in this study (color lines). The data is shown as the empty symbols, with 1$\sigma$ error-bars. The synthetic photometry from the model spectrum is shown with a cross symbol, due to the band integration, this does not always sit exactly on the spectrum. The location of the measurement is marked by a cross in Figure~\ref{FigData}.}
    \label{FigFitsPanel}
\end{figure*}

We investigate the differences in some of the key parameters from dust emission modeling, when the models presented here are all used in an identical fitting framework.
All residuals in this analysis are presented as \textit{(Data-Model)/Model}. For radial profiles, we use the pixel size as the annulus width and cover from the center of M101 to \rtf. 
Appendix~\ref{AppPrms} shows the resolved maps of the fitted parameters for each model. Appendix~\ref{AppResd} shows the residual maps of each model.

\subsection{Quality of fits}
\label{SecResiduals}
\begin{figure*}
    \centering
    \includegraphics[width=\textwidth]{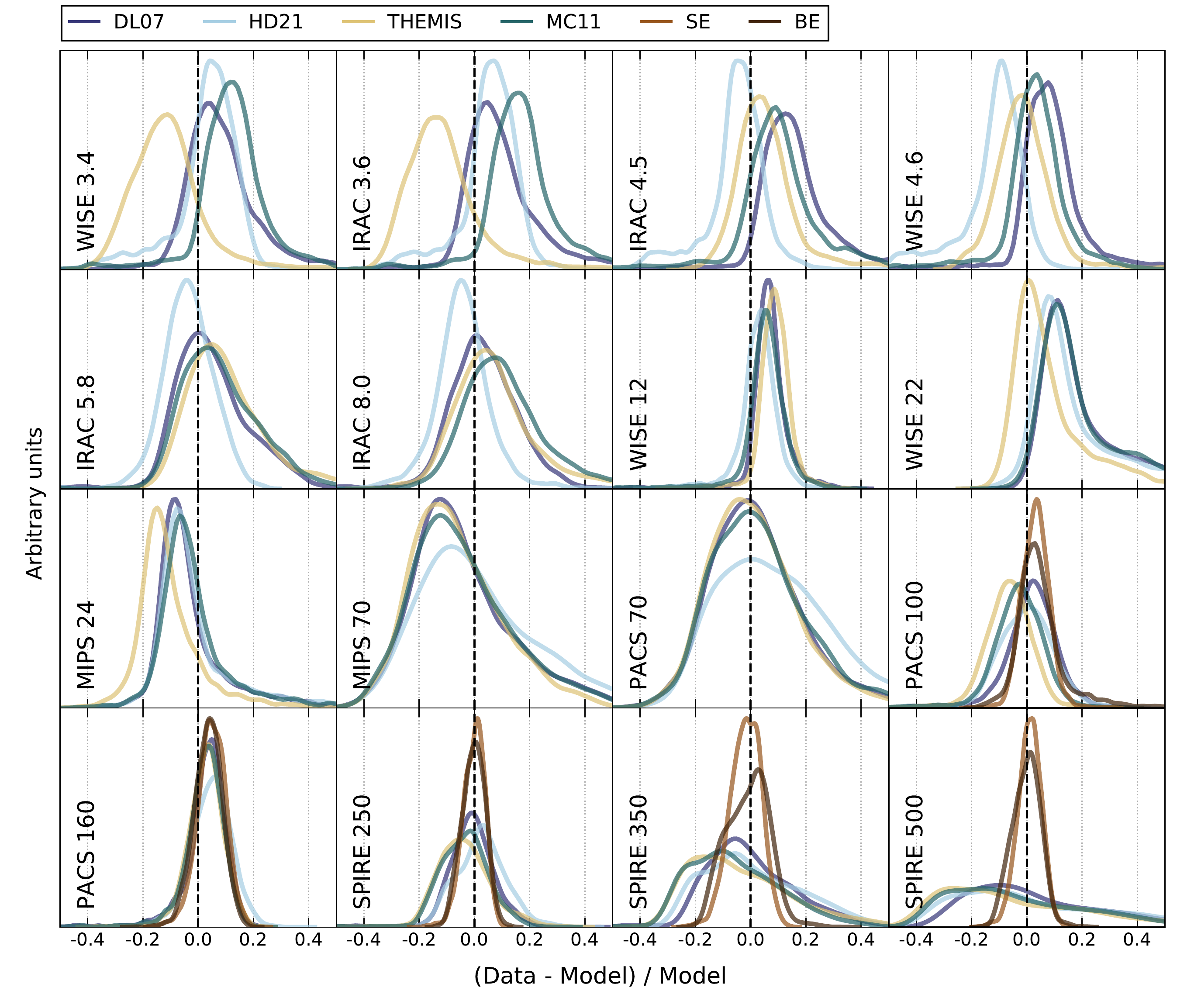}
    \caption{Fractional residuals \textit{(Data-Model)/Model}\ for each model in each band. We plot the (Gaussian) kernel density estimates of the the residual distributions.
    The WISE~12 band shows narrow, offset fits for all models while other mid-IR bands show clear over/under-estimations by some models.
    The physical models perform a good fit at 250~$\mu$m that gets progressively worse towards longer wavelengths. Only the modified blackbody models show systematically good fits, within 10\%, in all far-IR bands, likely because of the spectral index, $\beta$ being a free parameter.}
    \label{FigResiduals}
\end{figure*}

Figure~\ref{FigFitsPanel} shows an example of the fits for each model in a single pixel (marked by the cross in Figure~\ref{FigData}). 
In each panel we plot the best fit spectrum (colored lines) to the data SED (empty circles). The residuals are shown in the colored symbols. Negative residuals mean that the model overestimates the data. For example, the negative (and decreasing) residuals from 250 to 500~$\mu$m in the physical models panels are representative of a systematic overestimation of the data (present also in other locations; see Appendix Figures~\ref{AppResdDL}--\ref{AppResdHD}).
In Figure~\ref{FigResiduals}, we show the fractional residuals {\it (Data-Model)/Model}\: in each band, for the pixels above the 3$\sigma$ threshold. Appendix~\ref{AppMaxLkd} shows the reduced $\chi^2$ for all models.

The bulk of the residuals in the short wavelength bands are within the instrument uncertainties and calibration errors. For example, despite larger uncertainty due to the extended source correction, the IRAC~8 band shows residuals mostly within 10\%. 
Below 4~$\mu$m, \citetalias{THEMIS} shows a clear offset that may be related to the absence, or low amount, of an ionized component in the HAC population, which leads to enhance the mid-IR features. It is also worth noting these bands are dominated by starlight, modeled by a 5,000~K blackbody, which is independent from the dust model itself. 
The residuals at 12~$\mu$m are systematically positively offset by less than 10\%, but all physical models show very narrow residual distributions. This is in contrast with the broader residuals in IRAC~8 and WISE~22 bands, where we can see more differences between models.

All models show differences in the central values of the residual distribution at 100~$\mu$m. At 160~$\mu$m, all residuals overlap and models perform fits of similar quality. 
At longer wavelengths, significant difference begin to appear.
At all far-IR wavelengths, the modified blackbody models reproduce the SEDs the best. This is likely because of the additional parameter that can adjust the far-IR slope ($\beta$ in the \SE model and $\beta_2$ in the \BE model). 
The SPIRE~250 band shows symmetrical residuals centered on 0 for the physical models (except \citetalias[][]{THEMIS}), while the residuals get progressively worse at 350 and 500~$\mu$m for all physical models, showing that on average, the modeled far-IR slope of the SEDs is steeper than the data.

In the SPIRE~350 and SPIRE~500 bands, the large number of pixels underestimated by the models show residuals much larger than the uncertainties, ruling out statistical noise and indicating that the models are not able to fit these wavelengths.
In the SPIRE~500 band, some of the pixels show the so-called ``sub-millimeter excess'' seen in other studies \citep[e.g.][]{Galametz14, Gordon14, Paradis19}.

\subsection{Total Dust Mass and Average Radiation Field}
\label{SecTotalValues}
\begin{figure}
    \centering
    \includegraphics[width=0.5\textwidth, trim={0 0 2cm 0}, clip]{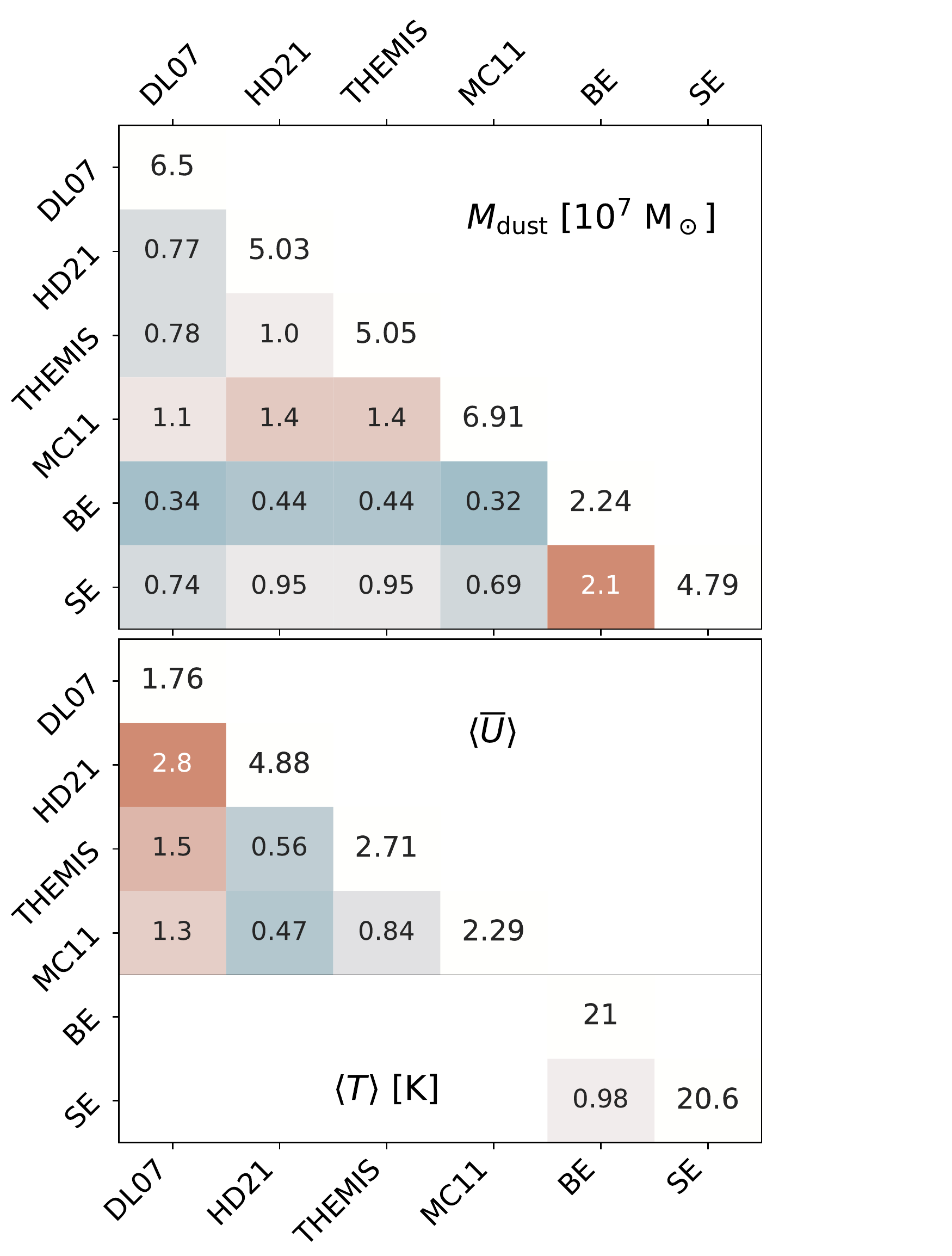}
    \caption{Integrated values (over the 3$\sigma$ pixels). The diagonal elements show the total dust mass (top panel), and radiation field (bottom panel, above the line) or temperature (bottom panel, below the line). The off-diagonal elements are ratios of the integrated values between different models, and are read as ``model Y-axis/model X-axis'' (e.g. $M_{\rm dust}^{\rm HD21} / M_{\rm dust}^{\rm DL07} = 0.77$).}
    \label{FigIntegRatios}
\end{figure}

We compute several galaxy averaged quantities: the total dust mass, $M_{\rm dust}$, the dust mass-weighted average radiation field \avubar for the physical models, and the mass-weighted average dust temperature $\langle T_{\rm dust} \rangle$ for the blackbody models. 
The average radiation field \ubar is calculated for each pixel as
\begin{equation}
    \overline{U} = (1-\gamma)\ U_{\rm min} + 
    \gamma \times U_{\rm min} \ \frac{{\rm ln(U_{max}}/U_{\rm min})}{1 - U_{\rm min}/{\rm U_{max}} }
\label{EquUbar}
\end{equation}
since we fixed $\alpha=2$ \citep[][]{Aniano20}. The galaxy-integrated \emph{mass-weighted} averages are calculated as
\begin{equation}
\begin{split}
    \langle \overline{U} \rangle &= \frac{\sum_j{\overline{U}_j \times \Sigma_{{\rm dust, }j}}}{\sum_j \Sigma_{{\rm dust, }j}} \\
    \langle T_{\rm dust} \rangle &= \frac{\sum_j{T_{{\rm dust, }j} \times \Sigma_{{\rm dust, }j}}}{\sum_j\Sigma_{{\rm dust, }j}}.
    \end{split}
\label{EquAvUbarT}
\end{equation}
The integrated values are calculated over the pixels above the 3$\sigma$ detection threshold.
In Figure~\ref{FigIntegRatios}, we show these measurements for each model as diagonal elements.
We also provide the ratios between all models in $M_{\rm dust}$ and $\langle \overline{U} \rangle$ (or $\langle T_{\rm dust} \rangle$) to explicitly show their differences.
These off-diagonal elements read as Y-model / X-model (e.g. $M_{\rm dust}^{\rm HD21} / M_{\rm dust}^{\rm DL07} = 0.77$). 

The top panel of Figure~\ref{FigIntegRatios} shows the total dust masses.
The \BE modified blackbody model yields the lowest total dust mass, while the \citetalias{Compiegne11} model yields the highest. The \SE model shows very different spatial variation of dust surface density than the other models (see Section~\ref{SecSigd}).
In a recent study of the KINGFISH sample, \citet[][]{Aniano20} found a total dust mass of $9.14 \times 10^7$~\msol (before their renormalization) by fitting the \citetalias{DL07} at MIPS~160 resolution ($\sim 39''$)\footnote{The difference between \citet{Aniano20}'s dust mass and ours is due to the larger area used in the former for the total dust mass calculation.}.
Using the \texttt{CIGALE} SED fitting tool \citep[e.g. ][ and reference therein]{Boquien19} and  \citetalias{THEMIS}, \citet[][]{Nersesian19} found a total dust mass of $4.70 \times 10^7$~\msol, which is similar to the $5.05 \times 10^7$~\msol from \citetalias[][]{THEMIS} in this study. It is worth noting that \citet[][]{Nersesian19} performed fits to the integrated SED, which could lead to a lower dust mass \citep[][]{Aniano12, Utomo19}. However, low signal-to-noise pixels are included in the integrated SED, and excluded in the resolved fits. 

It is interesting to note that despite the fairly good agreement ($\sim 10\%$, Figure~\ref{FigResiduals}) of all physical models with each other (and modified blackbody models at far-IR wavelengths) in reproducing the data, the differences in dust masses can be much larger. This suggests intrinsic opacity values rather than fit quality dominate the differences between models in dust mass.
This is supported by the recent, extensive study done by \citet[][]{Fanciullo20}. By comparing the literature opacities (including 3 models used in this work) with laboratory dust analogue opacities, they found that dust masses can be overestimated by more than an order of magnitude.

The bottom panel of Figure~\ref{FigIntegRatios} shows the integrated values for \avubar and $\langle T_{\rm dust} \rangle$.
As expected, the \citetalias{Hensley+Draine_2020c} model requires the highest radiation field, based on its colder dust. \citetalias{THEMIS} and the \citetalias{Compiegne11} model show similar values of \avubar.
\citet[][]{Nersesian19} found a dust temperature of $21.7$~K by fitting a modified blackbody (with the \citetalias{THEMIS} opacity), close to that yielded by the modified blackbody models here.

The mass-weighted average temperatures correspond to radiation fields of 2.5 and 2.3 (using Equ.~\ref{EquTpropU}) for the \BE and the \SE models, respectively. They are in good agreement with the radiation field values fitted by the physical models.

{}

\subsection{Dust Surface Density, \sigd}
\label{SecSigd}
\begin{figure*}
    \centering
    \includegraphics[width=\textwidth]{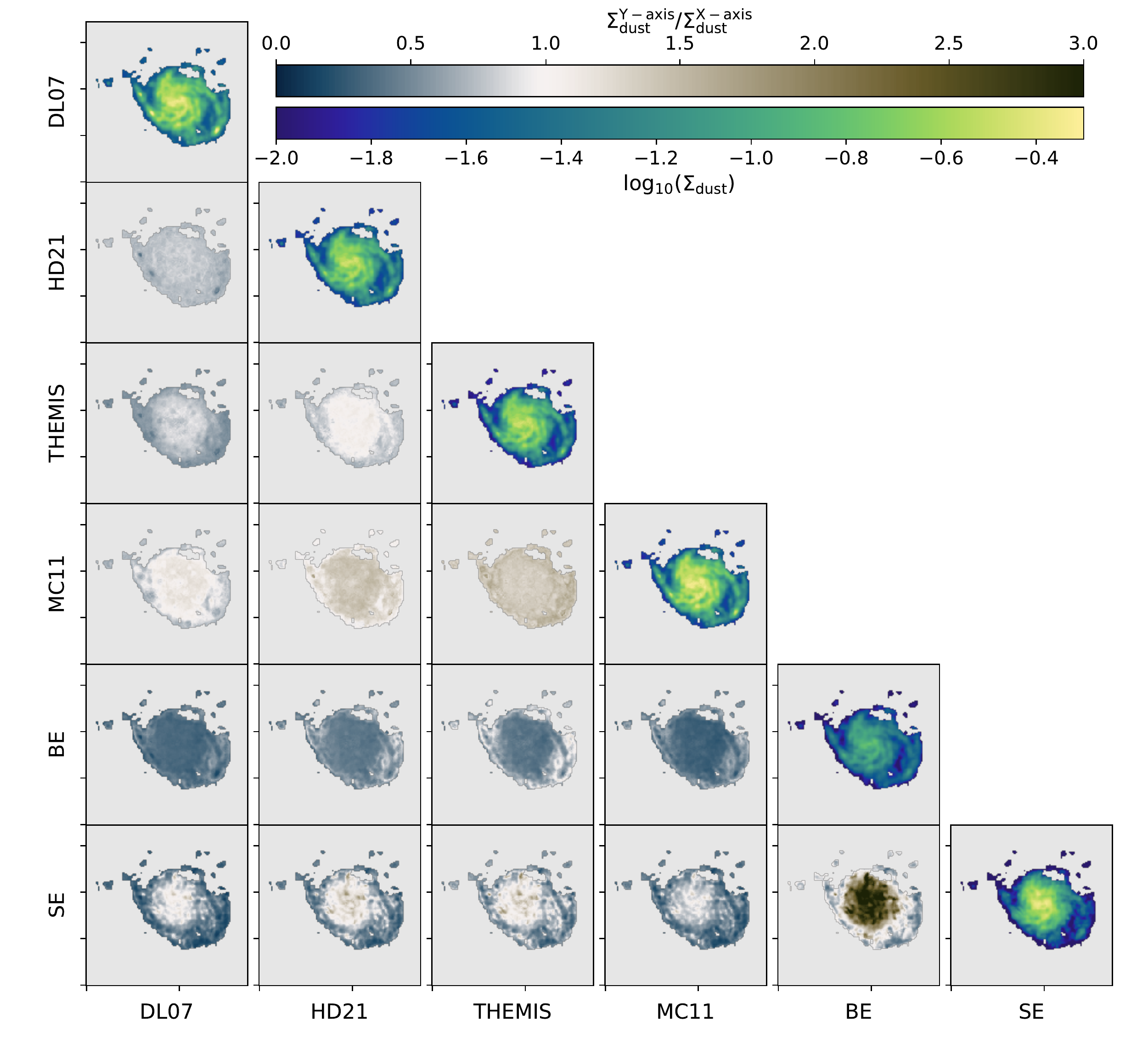}
    \caption{Dust surface density maps (diagonal) and corresponding ratios with each model. The \citetalias{Compiegne11} model requires the largest dust mass, followed by the \SE and \citetalias{DL07} models. All physical dust models show very similar spatial variations despite having different values of \sigd. The HD21/DL07 and MC11/THEMIS ratio maps are particularly smooth across the disk, but the MC11/DL07 and THEMIS/HD21 have the ratios closest to 1. The \SE model shows clear structural differences with the other models, by requiring a lot of dust in the center, but rapidly dropping in the outskirts.}
    \label{FigSigdRatios}
\end{figure*}

Figure~\ref{FigSigdRatios} shows maps of the dust surface density, \sigd, for each model, as well as their ratios with each other. 
We can see that the physical dust models \citetalias{DL07}, \citetalias{Hensley+Draine_2020c}, \citetalias{THEMIS} and \citetalias{Compiegne11} are all fairly close to each other (light colors), with variations in dust surface density within a factor of 2. They all yield similar dust surface density structures, and appear to vary from one another by a spatially smooth offset. 
\citetalias{THEMIS} and the \citetalias{Hensley+Draine_2020c} model show the closest \sigd values, but show an inversion of their ratio around 0.38~$r/$\rtf, where \citetalias{THEMIS} requires less dust (see Figure~\ref{FigRadProfSigd}). 
The HD21/DL07 and MC11/THEMIS ratio maps are particularly flat, with ratios $\sim 1.3$--$1.4$ in both cases, pointing at the resemblance in their large grains properties and size distribution. 
The \citetalias{Compiegne11} model requires the most dust mass. This is consistent with the comparison analysis in \citet[][]{Chastenet17}. 
The dust surface density from the \BE model is consistently lower than the physical dust models. It shows a rather smooth offset, which indicates that the spatial variations are fairly similar between them.
The dust surface density from the \SE model shows different spatial structures in the center and the outskirts of the galaxy. It yields high \sigd values in the center, but drops more rapidly than any other model with increasing radius \citep[Figure~\ref{FigRadProfSigd}; also][]{Chiang18}. 

A caveat of our approach is the difference in treating the heating of dust grains between physical dust models and modified blackbodies. In the latter, we only use a single temperature, which is not equivalent to the ensemble of radiation fields used in the physical models. However, given the two extreme behaviors of the modified blackbodies (the \SE model requiring a high dust mass, and the \BE model requiring the lowest), it is not obvious that the use of multiple temperatures (or radiation fields) drive the differences in dust mass observed here. 
Rather, the calibration of the modified blackbodies, and their effective opacity, seem to be more important.

{}

\subsection{Average radiation field, \ubar}
\begin{figure}
    \centering
    \includegraphics[width=0.5\textwidth]{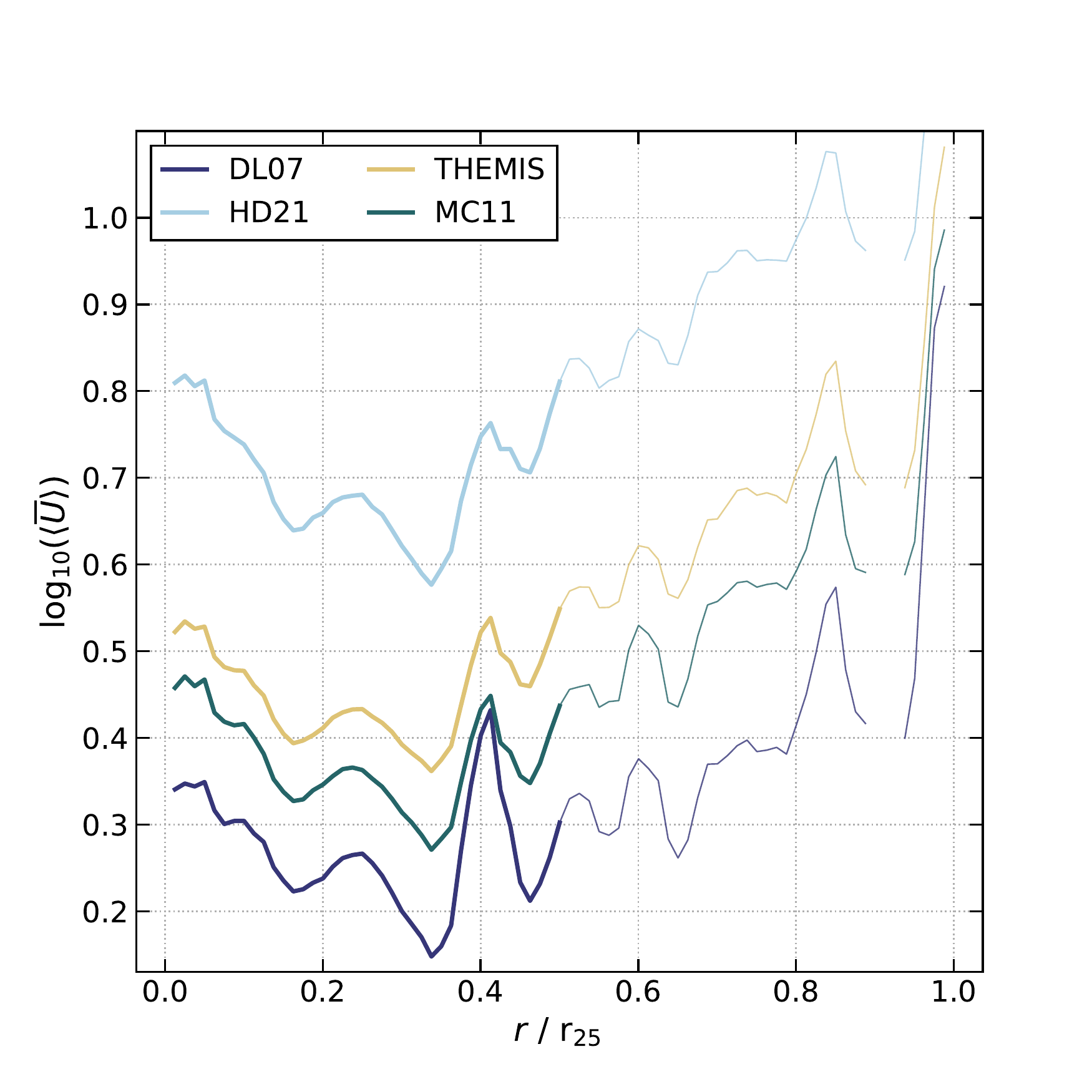}
    \hfill
    \includegraphics[width=0.5\textwidth]{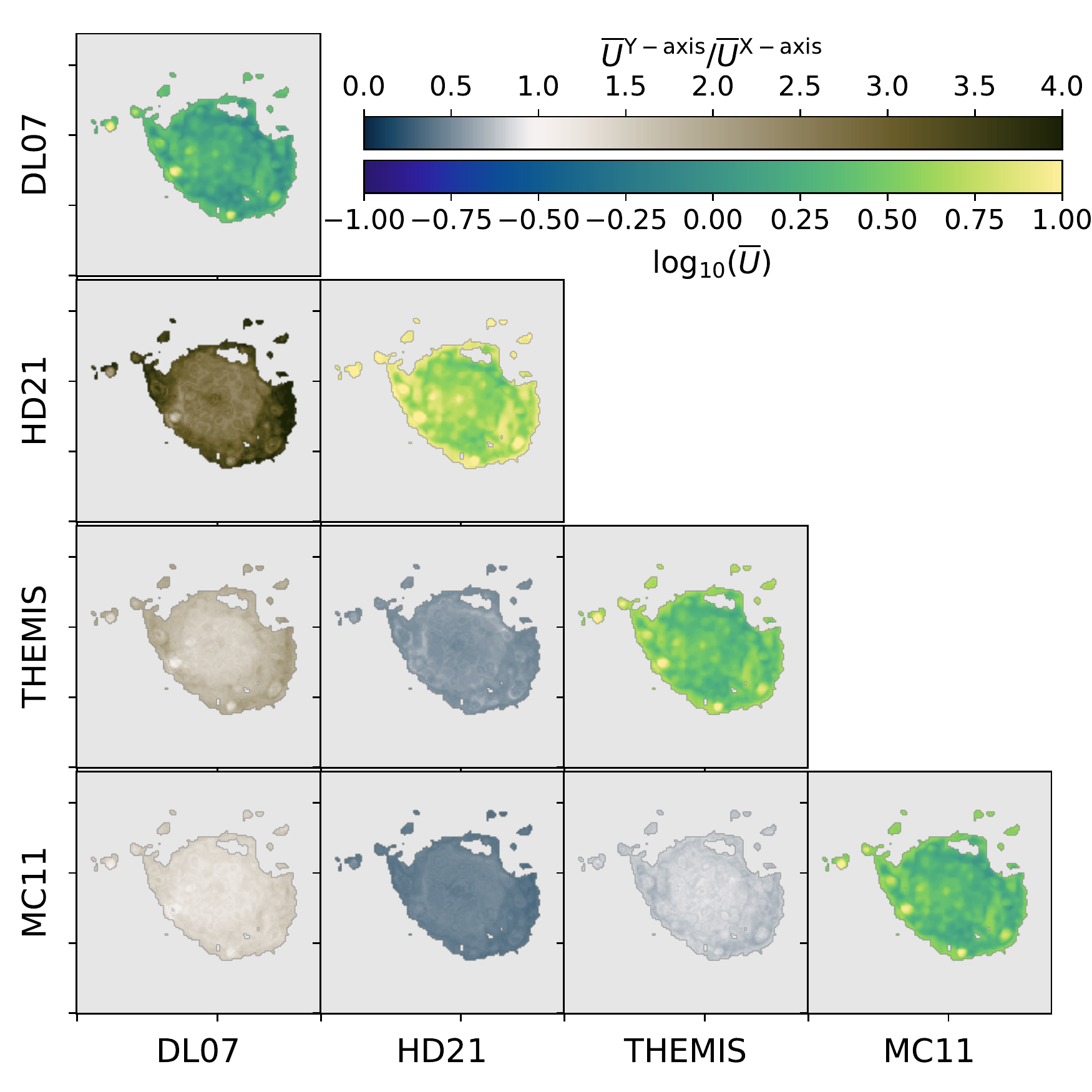}
    \caption{\textit{Top:} Radial profile of \avubar (\textit{mass} averaged radiation field, see Equ.~\ref{EquAvUbarT}) for each physical model. 
    The thick lines stop where the radial profile is affected by the selection effect due to fitting only bright pixels (Section~\ref{SecBkg}).
    \textit{Bottom:} Average radiation field, \ubar, maps (diagonal) and corresponding ratios with each model.
    The \citetalias{Hensley+Draine_2020c} model shows the highest values of \ubar in all the pixels. Despite different values, all models show very similar spatial distribution of \ubar, including in \ion{H}{2} regions.}
    \label{FigUminRatios}
\end{figure}

We perform the same ratio analysis with \ubar, derived from the fitted parameters in the physical dust models (Equ.~\ref{EquUbar} and \ref{EquAvUbarT}).
In Figure~\ref{FigUminRatios} we show the radial profiles for \ubar, as well as the parameter maps and the ratios of each model. 
In the radial profile, the thick lines stop where the selection effect due to fitting only bright pixels becomes important. Using the SPIRE~500 image, we found that radial profile of IR emission for all pixels and for fitted-pixels only differ significantly at $\sim 6'$ (i.e. 0.5~\rtf).
The variations in \ubar are reflective of those of \umin, since the $\gamma$ values are overall small, lending more power to the delta-function than the power-law (Equ.~\ref{EquRF}).

The overall variations of \ubar appear to be rather smooth, which is expected since it is dominated by the diffuse radiation field \umin.
However, we do find enhanced values of \ubar in \ion{H}{2} regions. The ratio maps do not strongly show these peaks, which indicates that all models behave similarly and require higher radiation field intensities in these regions.
Like for the \sigd parameter, the ratio maps for \ubar do not display any conspicuous spatial differences, but rather an offset between each model. This is also visible in the upper panel of Figure~\ref{FigUminRatios}.

The \citetalias{Hensley+Draine_2020c} model shows the highest values of \ubar. This is expected as the dust in the model is ``colder'' than other models, due to very few large carbonaceous grains. This leads to a higher radiation field intensity required to reach the same luminosity.

The spatial distributions of the $\gamma$ parameter are similar in all physical models, and we chose not to show them. Instead, we use the average radiation field, \ubar, that includes $\gamma$ in its calculation.
The \citetalias[][]{Hensley+Draine_2020c} model shows the lowest values of $\gamma$, which, combined to the highest values of \umin, means it requires more power in the delta function than the other physical models.

{}

\subsection{Fraction of PAHs}
Three models have an explicitly defined PAH fraction. The \citetalias{DL07} and \citetalias{Hensley+Draine_2020c} models define the parameter \qpah as the fraction of the dust mass contained in carbonaceous grains with less than 10$^3$ carbon atoms, roughly less than 1~nm in size. We use the $f_{\rm PAH}$ parameter from the \citetalias{Compiegne11} model to estimate a PAH fraction, i.e. mass of grains with sizes from 0.35 to 1.2~nm, as defined by the fiducial parameters. We refer to the mid-IR emission features carriers in \citetalias{THEMIS} as HACs.
We use the definition in \citet[][]{Lianou19} to compute a fraction of HACs from \citetalias{THEMIS} results, that can compare to the PAHs in other dust models: they found that this fraction of HACs corresponds to grains between 0.7 and 1.5~nm of the sCM20 component.
It is important to remember that the strict definition of the PAH/HAC fraction is different in each model, but its purpose---fitting the mid-IR emission features---remains similar.

\begin{figure}
    \centering
    \includegraphics[width=0.5\textwidth]{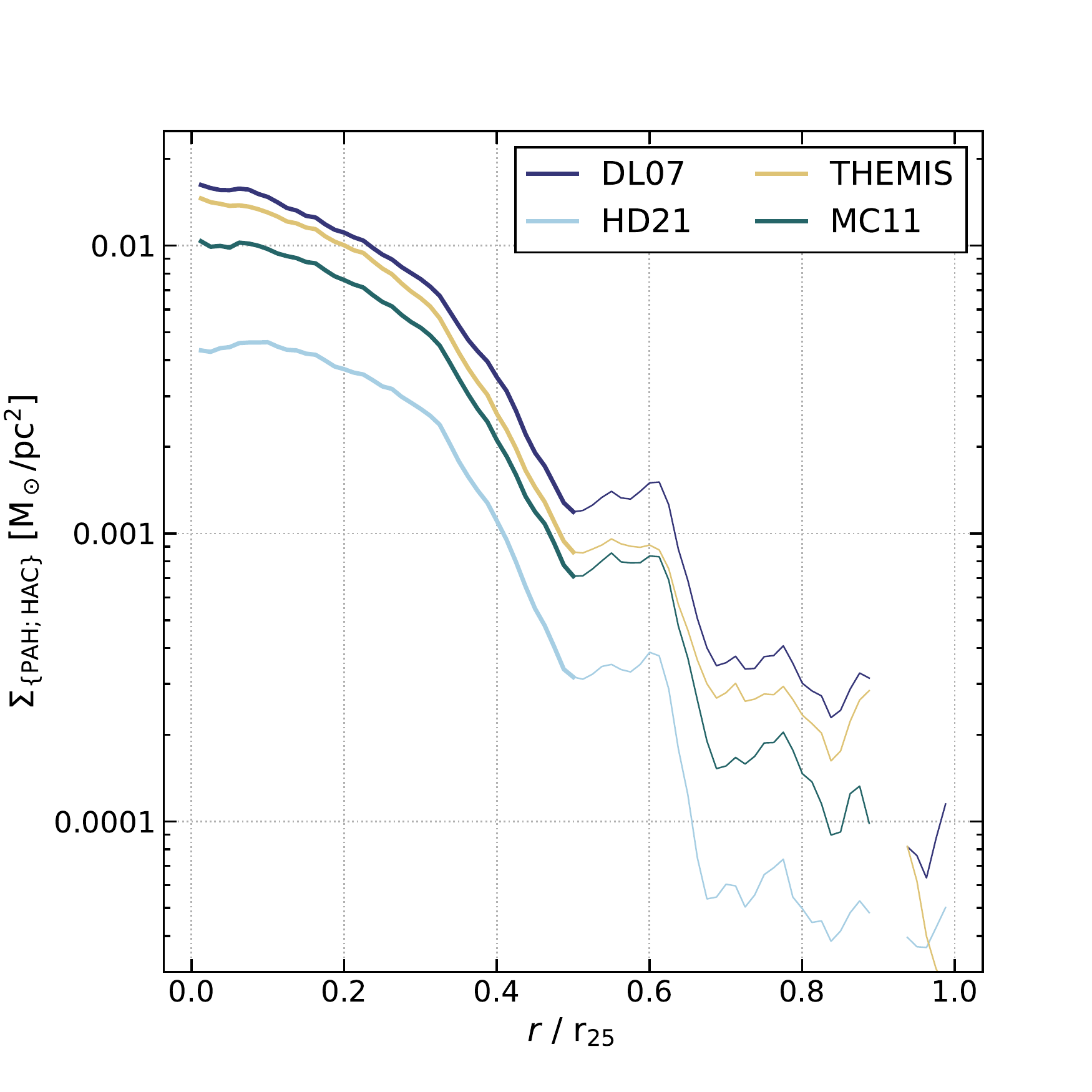}
    \hfill
    \includegraphics[width=0.5\textwidth]{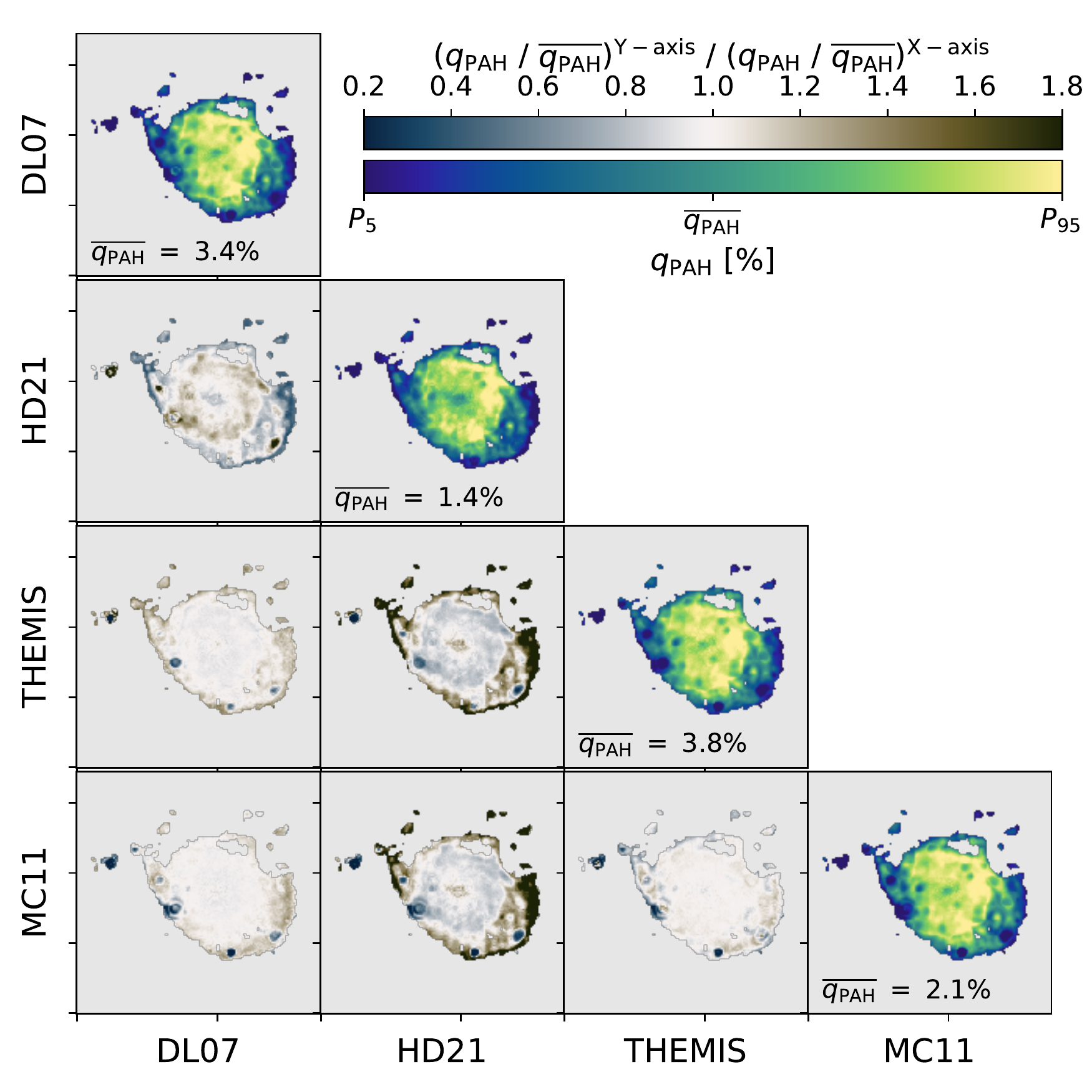}
    \caption{\textit{Top:} Radial profiles for $\Sigma_{\rm PAH}$, or $\Sigma_{\rm HAC}$, in \msolpc. The models yield very similar spatial variations. This indicates that the definition of the features carriers matters in terms of their contribution to the total dust mass, but they all reproduce the mid-IR features similarly.
    The thick lines stop where the radial profile is affected by the selection effect due to fitting only bright pixels (Section~\ref{SecBkg}).
    \textit{Bottom:} Fraction of PAHs (or HACs) (diagonal) centered on their respective mean value, with boundaries at 5 and 95 percentiles ($P_5$, $P_{95}$). The normalized ratios (normalized to the mean; off-diagonal) show the spatial variations between models. }
    \label{FigQPAHRatios}
\end{figure}

We investigate the variations of the surface density of the carriers, $\Sigma_{\rm PAH}$ and $\Sigma_{\rm HAC}$, instead of their abundances.
In the top panel of Figure~\ref{FigQPAHRatios}, we show the radial profiles of $\Sigma_{\rm \{PAH;~HAC\}}$. 
Although the absolute values of the surface density of PAHs (HACs) differ by a factor up to $\sim 3.5$ (similar to dust masses), their gradients are very similar. This behavior shows that the grain populations that are held responsible for the mid-IR features in each model follow comparable distributions. In these models, their contributions to the total dust mass vary significantly but all prove to be a good fit to the mid-IR bands (see also Figure~\ref{FigResiduals}).
This is also exemplified by the normalized ratio maps in Figure~\ref{FigQPAHRatios} (bottom panel). The dark colors in the outer-most pixels of the \citetalias[][]{Hensley+Draine_2020c} model are due to best \qpah fit consistent with 0\%.
To visualize the variations between models, we normalize each parameter map to their mean value (as shown in the color-bar labels). We are thus able to compare the spatial variations of the maps, and avoid the offsets due to the definition differences of PAHs or HACs.

The \qpah map in \citet[][using the \citetalias{DL07} model]{Aniano20} shows similar features to ours. A large portion of the disk of M101 has a rather constant distribution of \qpah, with conspicuous drops in \ion{H}{2} regions.  
In their study using the \citet*[][]{DBP90} dust model, \citet[][]{Relano20} found a flat radial profile of the small-to-large grain mass ratio, up to 0.8~\rtf ($\sim 9.1'$). Our maps of the fraction of PAHs, or HACs, present a somewhat flat distribution (variations less than 1\%) out to $\sim 0.3~$\rtf ($\sim 3.4'$; see Appendix~\ref{AppPrms}), and a steep change further out.

{}

\subsection{Reproducing the mid-IR emission features}
\label{SecMidIR}
\begin{figure*}
    \centering
    \includegraphics[width=\textwidth]{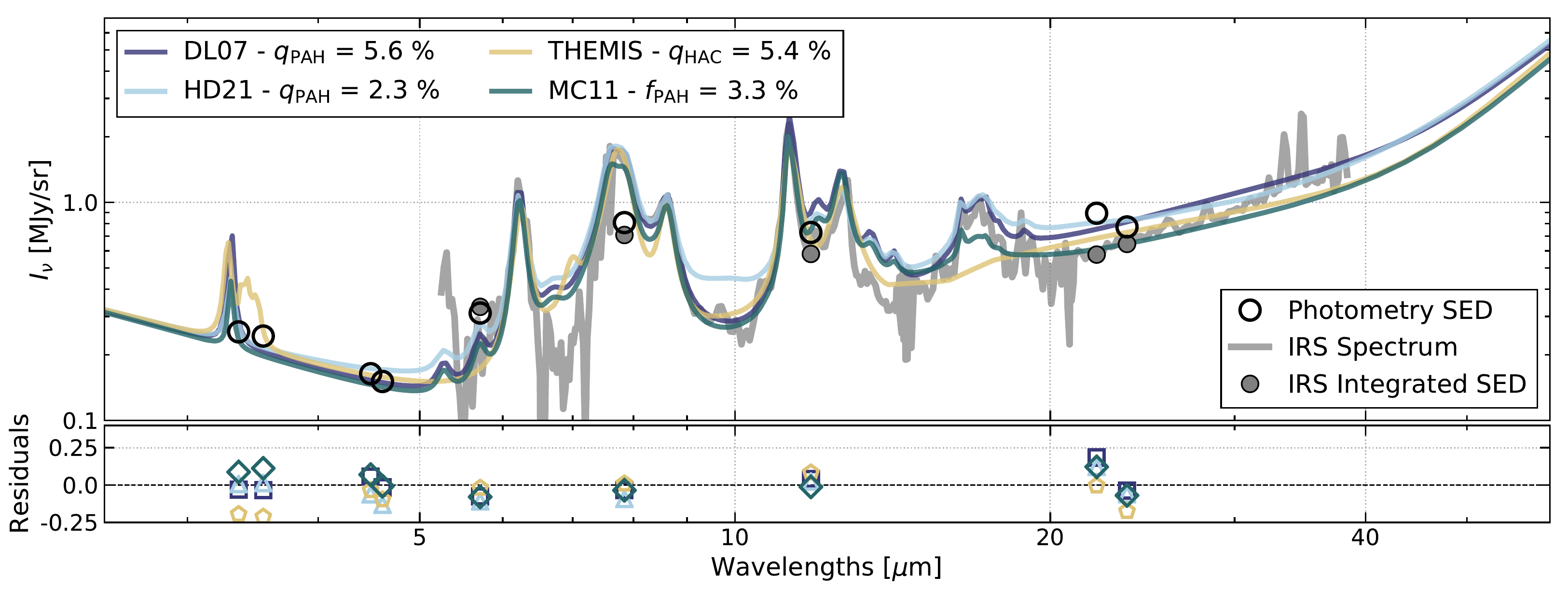}
    \caption{Fits to the integrated SED from the broad-band photometry (open circles) within a rectangle box (drawn in Figure~\ref{FigData}). The {\it Spitzer}/IRS spectrum and its corresponding SED (convolved in the 16 bands used here) are shown in gray (line and filled circles).
    All models perform a good fit to the mid-IR part of the SED, but show different fractions of PAHs (or HACs). Differences can be noticed between models: a higher 10~$\mu$m emission in the \citetalias[][]{Hensley+Draine_2020c} model, due to nano-silicates, or the lack of an emission feature at 17~$\mu$m in \citetalias[][]{THEMIS}.}
    \label{FigIRS}
\end{figure*}

To investigate in more detail the ability of each physical model to reproduce the PAH features, we perform a fit on an integrated SED and compare the results to measurements from the Infrared Spectrograph (IRS; on-board {\it Spitzer}) in that same region (J. D. T. Smith, private communication). 
Figure~\ref{FigIRS} shows a zoom on the mid-IR part of the models and the results of the fits to the integrated SED. From that fit, it appears that all models are able to generally reproduce the mid-IR features, with their different parameters, but we can notice a few differences between models.

From the residuals (bottom panel), we see that the models perform similarly at 5.8, 8.0 and 12~$\mu$m. 
At 22 and 24$\mu$m, the offset between the measurements means the models tend to split the difference and sit between the points. We note that all models appear to overestimate the continuum around 7~\micron, in between the 6.2 and 7.7~\micron\ PAH features. 

There are nonetheless a couple of noticeable differences between each model. For instance, the \citetalias[][]{Hensley+Draine_2020c} model shows a higher continuum at 10~$\mu$m than the other three models, despite a similar continuum at 20~$\mu$m. This rules out the higher \umin found in the \citetalias[][]{Hensley+Draine_2020c} model as the reason for the higher flux at 10~$\mu$m. Rather, in this model, the emission at $10~\mu$m is strongly dependent on the amount of nano-silicates, used to account for the lack of correlation between PAH emission and anomalous microwave emission \citep[][]{Hensley17}.
The ratio between the flux at 20 and 10~$\mu$m is $\sim 1.7$ for the \citetalias[][]{Hensley+Draine_2020c} models and 2 or above for the other three models.
On the other hand, \citetalias[][]{THEMIS} shows no emission feature around 17~$\mu$m, while the other models do (although it has no impact on this particular fit).
Around 7~$\mu$m, all models show a higher flux than the one seen in the IRS spectrum.

It is notable that, using only photometric bands from WISE and \spitzer/IRAC in the fit, all models reasonably well reproduce the mid-IR emission features, despite having different values of the PAH (or HAC) fraction and different definitions of the carriers. However, the comparison to spectroscopic measurements show that there are still differences between models.

{}

\subsection{SPIRE~500 and \sigd}
\label{SecSpireSigd}
\begin{figure*}
    \centering
    \includegraphics[width=\textwidth]{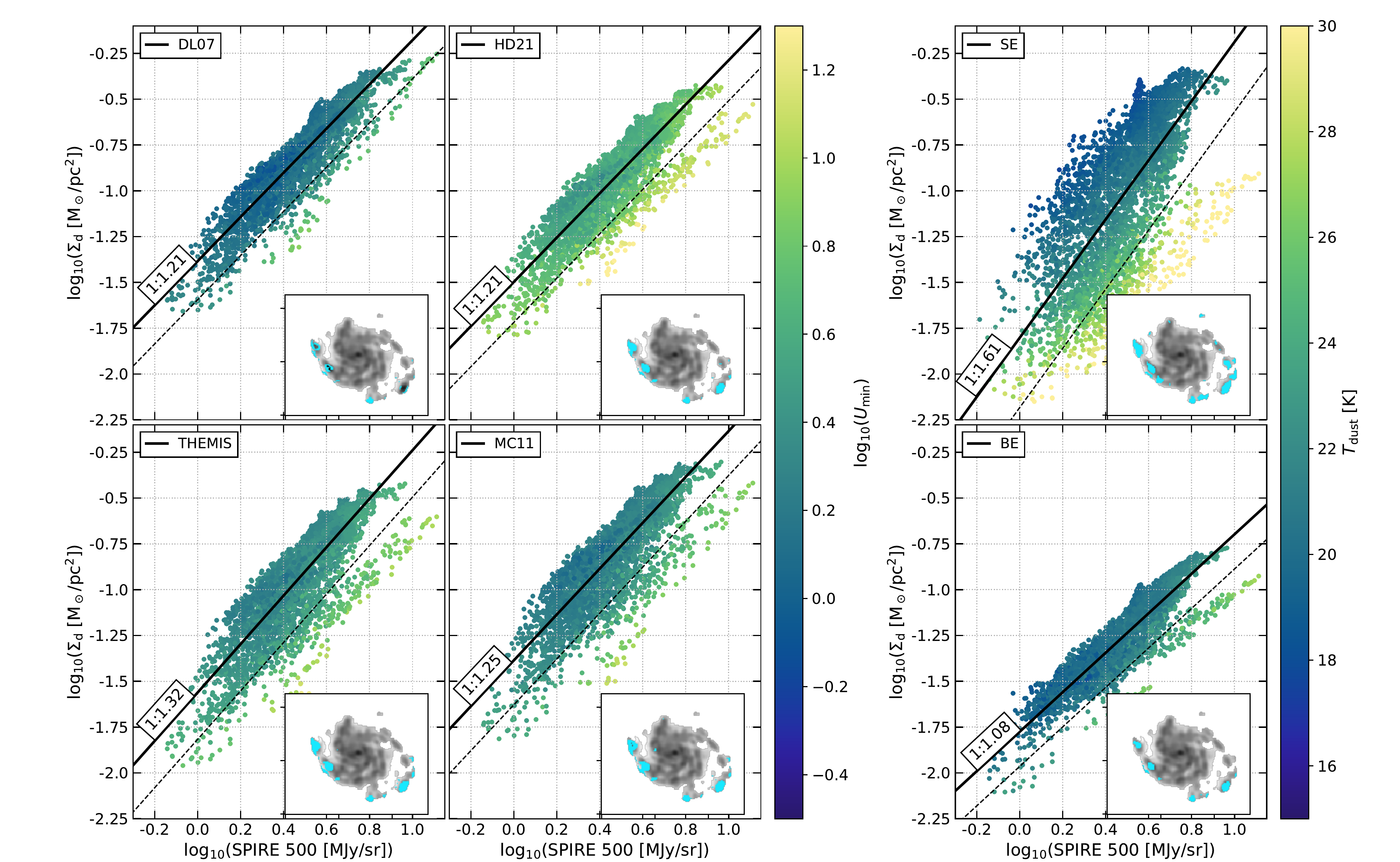}
    \caption{SPIRE~500 emission as a function of the fitted \sigd for each model, color-coded by the radiation field \umin or the dust temperature \td. We identify a separation in the linear scaling of \sigd with the dust emission at 500~$\mu$m, marked by the dashed lines. The pixels below the lines are plotted in color in the maps, and are located in \ion{H}{2} regions and surroundings. In these specific locations, the luminosity does not follow the same relation with \sigd than the rest of the galaxy.}
    \label{FigS500Sigd}
\end{figure*}

The monochromatic dust emission in far-IR wavelengths has often been used as a mass tracer of the ISM \citep[e.g.][]{Eales12, Berta16, Scoville17, Groves15, Aniano20, Baes20}.
In Figure~\ref{FigS500Sigd}, we plot the emission of M101 at 500~$\mu$m as a function of the fitted \sigd for each model (pixels above the 3$\sigma$ detection threshold), color-coded by the minimum radiation field \umin, or dust temperature \td.

In all cases, we can see two distinct relations as the SPIRE~500 emission increases. The majority of the fitted pixels show a linear scaling between the emission at 500~$\mu$m and \sigd, while in some specific regions of the galaxy, all models prefer a higher radiation field (or temperature) and a lower dust surface density. 
We provide the scaling relations between the emission at 500~$\mu$m in MJy/sr, and the dust surface density in \msolpc, for each model. 
We measure the 5$^{\rm th}$ and 95$^{\rm th}$ percentiles of the data points above the 3$\sigma$ detection threshold  (to keep the bulk of the distribution only). We fit a linear slope to these points\footnote{The fit coefficients and uncertainties were measured using the \texttt{numpy.polyfit} procedure.}:
\begin{equation}
\begin{split}
    &{\rm log_{10}(\Sigma_d)}^{\rm DL07} = 1.21\pm0.01 \times {\rm log_{10}}(I_\nu^{500\mu m})\\
    &\qquad \qquad \qquad - 1.38\pm0.005 \\
    &{\rm log_{10}(\Sigma_d)}^{\rm MC11} = 1.25\pm0.02 \times {\rm log_{10}}(I_\nu^{500\mu m})\\
    &\qquad \qquad \qquad - 1.39\pm0.008 \\
    &{\rm log_{10}(\Sigma_d)}^{\rm THEMIS} = 1.32\pm0.02 \times {\rm log_{10}}(I_\nu^{500\mu m})\\
    &\qquad \qquad \qquad - 1.56\pm0.008 \\
    &{\rm log_{10}(\Sigma_d)}^{\rm HD21} = 1.21\pm0.01 \times {\rm log_{10}}(I_\nu^{500\mu m})\\
    &\qquad \qquad \qquad - 1.50\pm0.006 \\
    &{\rm log_{10}(\Sigma_d)}^{\rm SE} = 1.61\pm0.03 \times {\rm log_{10}}(I_\nu^{500\mu m})\\
    &\qquad \qquad \qquad - 1.80\pm0.02 \\
    &{\rm log_{10}(\Sigma_d)}^{\rm BE} = 1.08\pm0.08 \times {\rm log_{10}}(I_\nu^{500\mu m})\\
    &\qquad \qquad \qquad - 1.78\pm0.005 .
\end{split}
\end{equation}

To identify the pixels that ``branch out'' from the bulk we select any pixel that falls below one standard deviation from the fit (dashed-lines). 
In each panel, we show the spatial location of these pixels. It becomes clear that the regions that need a higher \umin (or \td) are \ion{H}{2} regions and in the outskirts of the galaxy. These pixels can account between 4 and 11\% of the pixels above the 3$\sigma$ detection.
The branching-out from the main relation is likely the consequence of the fact that the dust in these \ion{H}{2} regions is significantly hotter than average (as shown by the enhanced \ubar in all models in Figure~\ref{FigUminRatios}).  Lower dust-to-gas ratios in the galaxy outskirts may also contribute to this trend. 

{}

In \citet[][]{Aniano20}, the authors found that this relation is well represented by a power-law scaling in the KINGFISH sample, with slope of 0.942, which is lower than our measured value of 1.2 for M101 with the \citetalias[][]{DL07} model. A linear slope would be expected if the dust temperature and optical properties were uniform throughout the galaxy, leading to a constant $I_{\nu}^{500}/\Sigma_{\rm dust}$ ratio. The spatial distribution of temperatures throughout each individual galaxy leads to distinct slopes, and for M101 we find that in regions of higher dust surface density, the dust is also warmer, leading to more 500~\micron\ emission (this can be seen in change in the color table on Figure~\ref{FigS500Sigd} at the highest $\Sigma_{\rm dust}$). The branch of \ion{H}{2} region points we see represents only a small fraction of the total data and may not be evident on the \citet{Aniano20} plot.

\section{Model Performance Given Abundance Constraints on Dust Mass}
\label{SecAnalysis}
\subsection{Maximum Dust Surface Density}
\label{SecMaxDust}
The calibration of dust models involves a constraint on elements locked in grains (see Section~\ref{SecModeling}). This step relies on depletion measurements, which characterize the distribution of heavy elements between the gas and solid phases. The final amount of elements allowed in dust grains varies between different physical dust models. The final dust masses derived by each model vary as well, as discussed in Sections~\ref{SecTotalValues} and \ref{SecSigd}.

A way to assess the performance of dust models is to verify that the required dust mass does not exceed the available heavy element mass, as constrained by metallicity measurements \citep[e.g.][]{Gordon14, Chiang18}. 
We perform this test in M101 since its metallicity gradient has been thoroughly characterized \cite[e.g.][]{Zaritsky94, Moustakas10, Croxall16, Berg20, Skillman20}.

We estimate the dust mass surface density upper-limit by assuming all available metals are in dust and calculating the metal mass surface density from the metallicity gradient and observed gas mass surface density:
\begin{equation}
    \frac{M_{\rm dust}^{\rm max}}{M_{\rm gas}} = \frac{M_{\rm Z}}{M_{\rm gas}} ,
\end{equation}{}
where $M_{\rm gas}$ and $M_{\rm Z}$ are determined as follows.

The gas surface density is the sum of \ion{H}{1} and H$_2$ surface densities including a correction for the mass of He (we neglect the ionized gas contribution). The latter is built from CO emission, assuming two prescriptions for the CO-to-H$_2$ conversion (see Section~\ref{SecData}).
We include the MW \aco prescription as it is widely used (Equ.~\ref{EquMW}).
We also choose the \aco prescription from \cite*{Bolatto13}, which takes into account environmental variations of \aco with metallicity and surface density (Equ.~\ref{EquBWL}). 
We emphasize that the \sigd upper-limit in this section is  dependent on the choice of \aco to derive a gas surface density. This is particularly true in the central region of M101, where H$_2$ dominates \citep[e.g.][]{Schruba11, Vilchez19} and where the two \aco differ the most.
Note also that this result differs with that of \citet[][]{Chiang18}. This is expected as the \aco conversion factor in their study \citep[from ][]{Sandstrom13} is lower than the ones used in this study, which leads to a lower upper-limit.

We use the $12+{\rm log_{10}(O/H)}$ radial profile from \citet[][]{Berg20} and convert it to metallicity through:
\begin{equation}
    \frac{M_{\rm Z}}{M_{\rm gas}} =     \frac{\frac{\rm m_ O}{\rm m_ H}\ 10^{ \big ( {\rm 12+log_{10}(O/H) \big ) -12}}}{1.36\ \frac{M_{\rm O}}{M_Z}},
\label{EquZtoG}
\end{equation}{}

\noindent with ${\rm m_O, m_H}$ the atomic masses of oxygen and hydrogen, respectively; and the oxygen-to-metals mass ratio $M_{\rm O}/{M_Z}=0.445$  \citep[][]{Asplund09}. 

\begin{figure*}
    \centering
    \includegraphics[width=\textwidth]{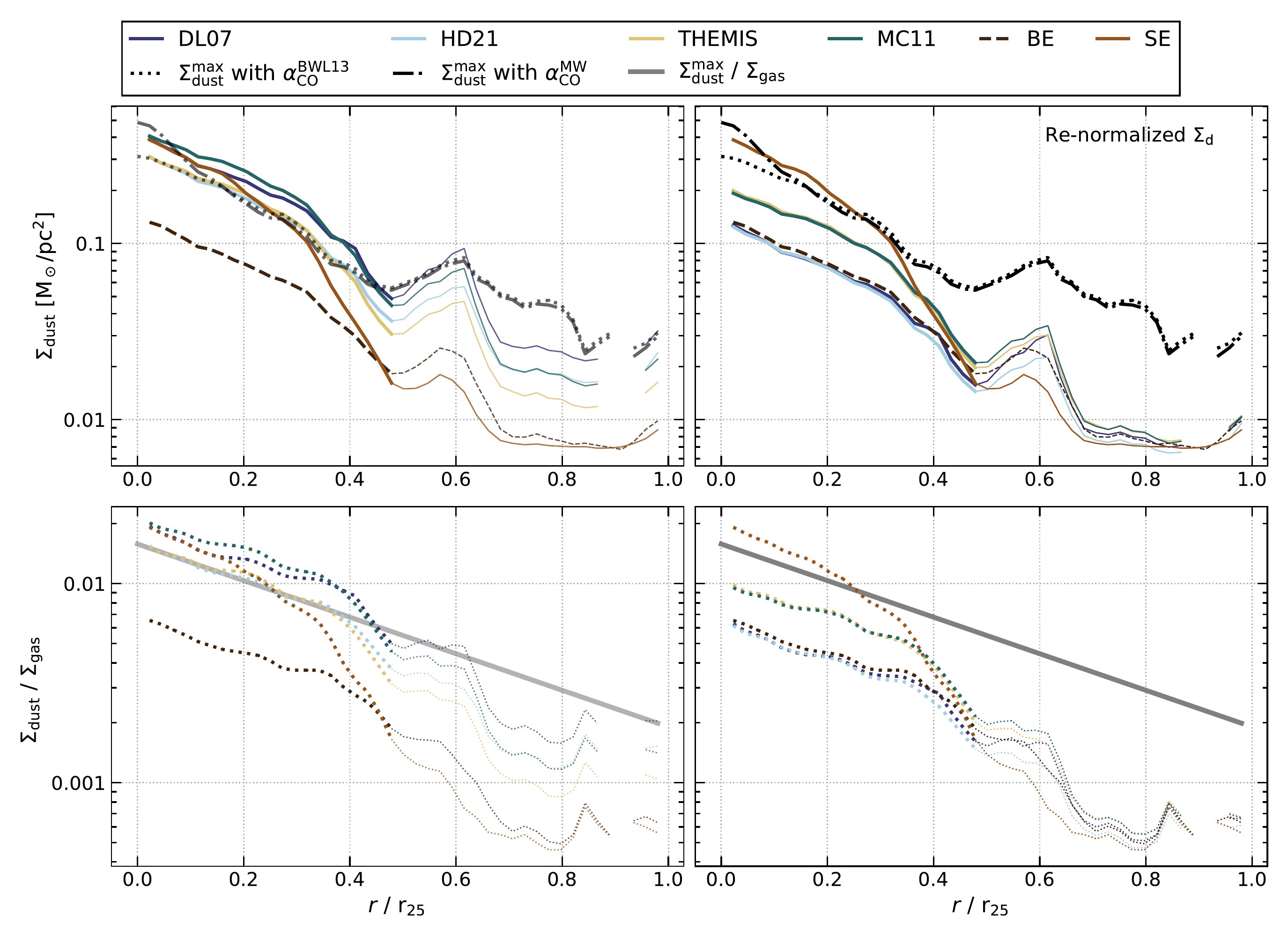}
    \caption{\textit{Top:} Radial profiles for \sigd (colored lines), and dust surface density upper-limits (black dotted and dash-dotted lines). Upper-limits are estimated from gas and metallicity measurements, assuming all metals are locked in dust grains (Section~\ref{SecMaxDust}). We have projected the dust surface density maps to the gas maps pixel grid, and masked the gas maps where there is no dust data (to ensure we are selecting identical pixels in building radial profiles).
    The upper-limits are invariant between the left and right panels.
    \textit{Left:} radial profiles from the fits. All physical models and the \SE model are either above or similar to the upper-limits, out to 0.4~$r$/\rtf.
    \textit{Right:} renormalized dust surface densities for the physical models (Section~\ref{SecRenorm}). The renormalization forces physical models to the same abundance constraints ($M_{\rm dust}/M_{\rm H}~=~1/150$) and to fit the same diffuse MW IR emission. Doing so, we derive correction factors and apply them to the dust surface densities, scaling them down to plausible values (below the $\Sigma_{\rm dust}^{\rm max}$ lines).
    \textit{Bottom:} Dust-to-gas ratios for each model, assuming the gas surface density derived with \aco from \cite*{Bolatto13}. The gray line represents the upper-limit from \citet[][]{Berg20} and assuming a dust-to-metal ratio of 1 (Equ.~\ref{EquZtoG}).
    The thick lines stop where the radial profile is affected by the selection effect due to fitting only bright pixels (Section~\ref{SecBkg}), creating the conspicuous features in the \ion{H}{2} region locations. The main bump at 0.6~$r/$\rtf corresponds to the two \ion{H}{2} regions NGC~5447 and NGC~5450.  The less visible bump at 0.5~$r/$\rtf corresponds to the \ion{H}{2} region NGC~5462.}
    \label{FigRadProfSigd}
\end{figure*}

The top panels in Figure~\ref{FigRadProfSigd} show the radial profile of \sigd for each model and the $\Sigma_{\rm dust}^{\rm max}$ upper-limits yielded by the two \aco prescriptions used here (black lines, dotted and dash-dotted).
We see on the top left panel that the \citetalias[][]{DL07} and \citetalias[][]{Compiegne11} models are above both $\Sigma_{\rm dust}^{\rm max}$ upper-limits in almost all the significant pixels. \citetalias[][]{THEMIS} and the \citetalias[][]{Hensley+Draine_2020c} models are fairly in line with the \citet*[][]{Bolatto13} upper-limit, with the conservative assumption that 100\% metals are in dust grains.
These behaviors suggest that the dust emissivity in all physical models is too low, which leads to requiring too much dust. We note that the large dust masses are likely due to the opacity calibration, rather than a wrong fit in the far-IR bands: in Figure~\ref{FigResiduals}, all physical models show a reasonable fit at 160~$\mu$m, much closer to the IR peak than the 500~$\mu$m band. We are confident that the IR peak is correctly recovered, and that the high dust masses are not due to the sub-millimeter excess.
Based on the reasonable quality of the fits, we believe that the excessive mass is likely due to the opacity calibration.

{}

\subsection{(Re-)Normalization}
\label{SecRenorm}
Despite sharing a common calibration approach, the details of the opacity calibration in the dust models used in this study vary in small but significant ways.
While all models were calibrated to MW diffuse emission, they did not use exactly the same high-latitude cirrus spectrum. 
In addition, the $M_{\rm dust}/M_{\rm H}$ adopted for the MW diffuse ISM by the physical dust models varies.
Additionally, the radiation field that best reproduces the MW diffuse emission, $U^{\rm MW}$, differs slightly from one model to the next. Because of the relationship between dust temperature and radiation field $(U \propto T_{\rm dust}^{4+\beta})$ and dust temperature and luminosity, even a slight difference in the assumed radiation field may lead to a significant change in the model's calibrated dust opacity.

To investigate calibration discrepancies, we re-normalize each of the dust models via a fit to a common MW diffuse emission spectrum using the same abundance constraints.
We use the MW SED described in \citet[][]{Gordon14}, which we previously used to calibrate $\kappa_\nu$ of the modified blackbody models \citep{Chiang18}. This SED is the same as that used in \citet[][]{Compiegne11}, a combination of DIRBE and FIRAS measurements \citep[e.g.][]{Boulanger96, Arendt98}\footnote{Although more recent measurements from \textit{Planck} are available in the far-IR, we emphasize here that the important aspect is about uniformity. We choose the DIRBE+FIRAS SED as it is conveniently the one used to calibrate the modified blackbody models. Additionally, the significant input brought by \textit{Planck} measurements are past the wavelength range used in our study (sub-milimeter and millimeter range).}. We do not use the ionized gas correction because depletions measurements do not correct for it, and instead use a correction factor of 0.97 for molecular gas only \citep[][]{Compiegne11}. 
We integrate the SED in the PACS~100, PACS~160, SPIRE~250, SPIRE~350 and SPIRE~500 bands, so all models use the same wavelength coverage. We use a 2.5\% uncorrelated and 5\% correlated errors to account for FIRAS and DIRBE uncertainties \citep[][]{Gordon14}.
The $M_{\rm dust}/M_{\rm H}$ ratio set in the normalization is 1/150, as suggested by depletions studies \citep[$F_*=0.36$ from][see also \citeauthor{Gordon14} \citeyear{Gordon14}]{Jenkins09}.

To perform the re-normalization using the MW SED, we use the same fitting technique as previously described with the following choices:
1) We do not use the combination of radiation fields, nor the stellar component (i.e. $\Omega_* = \gamma = 0$).
2) We allow the minimum radiation field \umin and the total dust surface density \sigd to vary in each physical model.  
3) We keep the relative ratios between grain populations fixed and do not vary them independently (e.g. for each model, we use the total spectra solid lines `DL07', `HD21', `MC11' and `THEMIS' in Figure~\ref{FigModels}). 

The fits yield renormalization factors that correct all physical models from their respective assumptions to a dust-to-H ratio of 1/150.
These corrections range from 1.5 for \citetalias[][]{THEMIS}, to 3 for the \citetalias[][]{DL07} model (see Appendix~\ref{AppModelTable}). With this normalization, we are able to meet the metallicity constraints.
The top right panel of Figure~\ref{FigRadProfSigd} shows the radial profiles with the correction factors applied to the surface densities of the physical dust models. The renormalization brings the models to lower dust surface densities that agree with the upper-limit based on the metal content.  It is interesting to note that the renormalized models now show three distinct behaviors in the dust mass surface density radial profile: \citetalias[][]{DL07}, \citetalias[][]{Hensley+Draine_2020c} and the broken power-law emissivity modified blackbody yield very similar results; \citetalias[][]{THEMIS} and \citetalias[][]{Compiegne11} are very similar to each other and offset by a factor of $\sim 2$ from the first group; and the simple power-law emissivity modified blackbody has a steeper increase that still puts it above the abundance constraints even though it is similarly normalized to the MW cirrus spectrum. 
The two different behaviors, for \citetalias{DL07} and \citetalias{Hensley+Draine_2020c}, and for \citetalias{THEMIS} and \citetalias{Compiegne11} are likely due to the difference in the best $U_{\rm min}^{\rm MW}$ to fit the MW SED, and their initial spectrum used for calibration.  

In the bottom panels of Figure~\ref{FigRadProfSigd}, we show the radial profile of the dust-to-gas ratios (DGR) for each model, using the $\alpha_{\rm CO}^{\rm BWL13}$ conversion factor. The bottom left shows the DGR with the derived dust surface densities, and the bottom right panels is after re-normalization. The thick gray line shows the upper-limit of the DGR using the metallicity gradient from \citet[][]{Berg20} and Equ.~\ref{EquZtoG}, with $\alpha_{\rm CO}^{\rm BWL13}$. We find the same abrupt change in the DGR as \citet[][]{Vilchez19} around 0.5~\rtf, although slightly lower values, consistent with the higher dust surface densities in this study.

\section{Discussion}
\label{SecDiscussion}
Here we discuss some next steps that should be undertaken to improve dust emission modeling.
In the previous sections, we investigated some systematic effects linked to the modeling choices.

We showed that physical dust models are likely to require too much dust mass, exceeding what is available based on metallicity measurements (for reasonable choices of CO-to-H$_2$ conversion factors, though we note that, in the central region, this assertion is strongly dependent on this choice).
This excess is linked to the calibration of these models, in particular the elemental abundances prescribed, and their assumed radiation field. 
Combined with the growing number of metallicity measurements in nearby galaxies \citep[e.g.][]{Kreckel19, Berg20}, additional constraints for external environments (beyond the MW) may help to perform better fits of dust emission. 

Several aspects of galaxy evolution studies rely on grain properties. 
Dust evolution models heavily rely on observational constraints to find the parameters that best match the observed properties of dust. 
The balance between destruction and formation processes in dust evolution models are adjusted by observed dust masses, that need to be accurately measured (e.g. by emission fitting).
Similarly, dust-to-gas ratio evolution with metallicity is often derived using dust masses from emission measurements \citep[e.g.][]{RemyRuyer14, DeVis19, Nersesian19, DeLooze20, Nanni20}, and are subject to the systematic biases found in this work.

Our study was designed for a rigorous comparison between models fit to mid- through far-IR SEDs.  Several choices made in this study are justified by our implementation of each model in the most similar framework possible. This also requires using a uniform radiation field description, and the parameters that go into describing the physical dust models in their fiducial form are not always adapted to the choice of the radiation field model of this work (Equ.~\ref{EquRF}).

Because of the limited wavelength coverage and SED sampling of this study, we ``tie'' together different grain populations, based on the similarity of their respective emission spectra.
In the \citetalias{Compiegne11} the large carbon (LamC) and silicate (aSil) grains have very similar slopes in the far-IR, which would make these fit parameters strongly degenerate if both were allowed to vary. 
In \citetalias{THEMIS}, the key difference between the large carbon grains (lCM20) and the large silicate grains (aSilM5) is their slopes in the far-IR: the lCM20 grains have a flatter SED than aSilM5. However, the spectral coverage used in this study is too limited to properly constrain the emission from the two grain populations.
These choices have implications for the evolution of dust composition in the ISM, since the ratios of carbon-to-silicate in large grains is assumed to be constant for each model of this study. 

Additionally, our further tests show that the $\gamma$ parameter is degenerate with the emission of some of the grain population in \citetalias{Compiegne11} and \citetalias{THEMIS}.
The abundance of small amorphous carbon (SamC) in \citetalias{Compiegne11} helps adjust the slope between 24 and 70~$\mu$m. In the radiation field parameterization chosen for this study, the $\gamma$ parameter has a similar impact on the shape of the dust emission. Keeping both the small amorphous carbon grains abundance and $\gamma$ as free parameters introduces a degeneracy in the fitting. For this reason we choose to keep the fiducial relative abundances of SamC with respect to that of big grains (LamC+aSil) fixed.
In \citetalias[][]{THEMIS}, when allowing both lCM20 and aSilM5 populations to vary (e.g. with  $f_{\rm aSil}$, the fraction of large grains in the form of aSilM5), we also introduce a degeneracy with the $\gamma$ parameter. 
A varying ratio of carbon-to-silicate grains performs a similar change of the SED shape and both parameters, $\gamma$ and $f_{\rm aSil}$ become slightly degenerate.  Future studies with more wavelength coverage and more detailed constraints on individual elemental abundances may be able to allow for more free parameters in the fits.

\section{Conclusions}
In this study, we compared the dust properties of M101 derived from six dust models: four physical dust models and two blackbody models. We used the models from \citet[][]{DL07}, \citet[][]{Compiegne11}, \citet[][THEMIS]{THEMIS} and \citet{Hensley+Draine_2020c}, as well as a \SE and a \BE modified blackbody models to assess the differences in various dust properties yielded by fitting the mid- to far-IR emission from {\it WISE}, the {\it Spitzer Space Telescope} and the {\it Herschel Space Observatory} photometry.
Our main conclusions are:
\begin{itemize}
\renewcommand\labelitemi{\tiny$\bullet$}
\setlength\itemsep{0.02cm}
    \item There are a few notable trends in the fitting residuals (described as \textit{(Data-Model)/Model}; Figure~\ref{FigResiduals}).
    All physical models reproduce the mid-IR bands within 10\%, with very similar residual distributions in the WISE~12 band.
    All models perform fits of similar quality at 160~$\mu$m. While the modified blackbody models can reproduce the data in all far-IR bands (residuals centered on 0), the fits from physical models have large residuals at long wavelengths.  This suggests that the flexibility to adjust the long wavelength slope of the opacity is important to reproduce the observed SEDs.
    \item All physical models reproduce the mid-IR emission features but yield different values of the mass fraction of their carriers (Figure~\ref{FigQPAHRatios}). Models that attribute the mid-IR emission features to PAHs or HACs do similarly well in reproducing the mid-IR spectrum. 
    \item We provide scaling relation of $\Sigma_{\rm dust} = f(I_\nu^{500~\mu m})$, and identified a diverging relation in \ion{H}{2} regions, where hot dust changes the relationship between dust emission and mass (Figure~\ref{FigS500Sigd}).
\end{itemize}{}
Examining the fitting results of total dust masses and dust surface density distributions, we find:
\begin{itemize}
\renewcommand\labelitemi{\tiny$\bullet$}
\setlength\itemsep{0.02cm}
    \item Models yield different total dust masses, up to a factor of 1.4 between physical models, and up to 3 including modified blackbodies (Figure~\ref{FigIntegRatios}), but all show similar spatial distributions of dust surface density (note the fairly low discrepancy between dust masses from physical models, compared to modified blackbody models). The \citetalias{Compiegne11} model requires the highest dust mass, and the \BE model the lowest.
    \item We use metallicity and gas measurements to calculate a dust surface density upper-limit (assuming all metals in dust) and show that all physical dust models require too much dust over some radial ranges in M101. Only the \BE modified blackbody model is below the upper limit of $\Sigma_{\rm dust}^{\rm max}$ (Figure~\ref{FigRadProfSigd}). This finding is dependent on the chosen prescription for the CO-to-H$_2$ conversion factor.
    \item To investigate the differences between dust masses and their relationship to the available heavy elements, we renormalized the models via fits to the same SED of the MW diffuse emission, assuming a strict abundance constraint of $M_{\rm dust}/M_{\rm H} = 1/150$ (Section~\ref{SecRenorm}). We derive scaling factors and apply them to the fitted dust surface density, and find renormalized dust mass values lower than $\Sigma_{\rm dust}^{\rm max}$ (Figure~\ref{FigRadProfSigd}). We find that the choices made to calibrate dust models have a non-negligible impact on the derived dust masses. 
\end{itemize}{}

To provide the strictest comparison, we do not always use dust models in their fiducial aspect, sometimes assuming a fixed ratio between two dust grain populations. 
The observational constraints brought by IR emission fitting are used to validate evolution models or derive scaling relations like the dust-to-gas ratio.
Our results show that these derived dust properties have systematic uncertainties that should be taken into account. 
Although there are still systematic uncertainties inherent in \ion{H}{2} region metallicity measurements, resolved metallicity gradients in nearby galaxies can be helpful for testing the opacity calibrations in dust models.

\acknowledgements
We thank the referee for a very thorough reading and providing detailed comments about the manuscript, which greatly improved the clarity of the paper.
The work of JC, KS, IC, AKL, and DU is supported by NASA ADAP grants NNX16AF48G and NNX17AF39G and National Science Foundation grant No.~1615728. The work of AKL and DU is partially supported by the National Science Foundation under Grants No.~1615105, 1615109, and 1653300.
TGW acknowledges funding from the European Research Council (ERC) under the European Union’s Horizon 2020 research and innovation programme (grant agreement No. 694343).

This work uses observations made with ESA \textit{Herschel} Space Observatory. \textit{Herschel} is an ESA space observatory with science instruments provided by European-led Principal Investigator consortia and with important participation from NASA. 
This publication makes use of data products from the Wide-field Infrared Survey Explorer, which is a joint project of the University of California, Los Angeles, and the Jet Propulsion Laboratory/California Institute of Technology, funded by the National Aeronautics and Space Administration. 
This work is based in part on observations made with the Spitzer Space Telescope, which was operated by the Jet Propulsion Laboratory, California Institute of Technology under a contract with NASA.
This research made use of \texttt{matplotlib}, a Python library for publication quality graphics \citep{Hunter:2007}.
This research made use of Astropy, a community-developed core Python package for Astronomy \citep{2018AJ....156..123A, 2013A&A...558A..33A}.
This research made use of NumPy \citep{van2011numpy}.
This research made use of SciPy \citep{Virtanen_2020}.
This research made use of APLpy, an open-source plotting package for Python \citep[][]{aplpy12, aplpy19}.
We acknowledge the usage of the HyperLeda database (\url{http://leda.univ-lyon1.fr}).

\bibliography{benchmark}{}
\bibliographystyle{aasjournal}

\appendix
\section{Calibration Details}
\label{AppModelTable}
We present here the details of the calibration methodology used in each physical model, and a summary of the calibration constraints.

\subsection{\citet[][]{DL07}}
\label{AppDL07}
\citetalias{DL07} was calibrated using the following constraints. 
The extinction is described in \citet[][]{WD01Ext} and uses the \citet[][]{Fitzpatrick99} extinction curves with a normalization of N$_H/E(B-V) = 5.8 \times 10^{21}$ H/cm$^2$ or $A_V/N_H = 5.3\times 10^{-22}$ cm$^2$.
The high-latitude cirrus emission per H observed by DIRBE (Diffuse Infrared Background Experiment) and FIRAS \citep[Far Infrared Absolute Spectrophotometer;][]{Arendt98, Finkbeiner99} is used as a reference for the far-IR emission, complemented by mid- and near-IR emission from IRTS \citep[Infrared Telescope in Space;][]{Onaka96, Tanaka96}.
\citet[][]{Weingartner01} adopt solar abundances from \citet[][]{Grevesse98}, assuming 30\% of carbon is in the gas phase. They assume all silicon is depleted and has abundance equal to the Solar value. \citetalias{DL07} uses $M_{\rm dust}/M_{\rm H} = 1.0\times 10^{-2}$.
The radiation field used in the \citet[][]{DL07} model is based on \citet[][]{Mathis83}.

\subsection{\citet[][]{Compiegne11}}
\label{AppMC11}
\citetalias{Compiegne11} was calibrated using the following constraints.
Extinction constraints were taken from \citet[][]{Mathis90} and \citet[][]{Fitzpatrick99}, including the $R_V = 3.1$ extinction curve in the UV-visible and a normalization of $N_H/E(B-V) = 5.8\times 10^{21}$ H/cm$^2$.
At $\lambda >25~\micron$, \citetalias{Compiegne11} use the MW cirrus emission per H observed by COBE-DIRBE and WMAP \citepalias[integrated in the \herschel and Planck/HFI bands; see][]{Compiegne11}. At $\lambda \leq 25~\micron$, a compilation of mid-IR observations of high latitude MW cirrus are used (combining measurements from AROME, DIRBE, and ISOCAM; we refer to reader to the \citeauthor{Compiegne11} paper for details). They scale the emission SED by 0.77 to account for ionized and molecular gas not accounted in the H column.
The allowed dust-phase abundances for C, O, and other dust components come from the difference between Solar (or F/G star) abundances and the observed gas phase abundances. In total, the $M_{\rm dust}$/$M_{\rm H} = 1.02\times 10^{-2}$.
\citetalias{Compiegne11} assumes the \citet[][$D_{\rm G} = 10$ kpc]{Mathis83} solar neighborhood radiation field to heat the dust grains.

\subsection{THEMIS}
\label{AppTHEMIS}
\citetalias{THEMIS} was calibrated using the same constraints as \citet[][]{Compiegne11} presented in the previous Section, with the addition of the far-IR-to-extinction relation $\tau_{250}/E(B-V)=5.8\times 10^{-4}~$ \citep[][]{PlanckCollabTHEMIS}.
In \citetalias{THEMIS}, $M_{\rm dust}/M_{\rm H} = 7.4\times 10^{-3}$.

\subsection{\citet[][]{Hensley+Draine_2020c}}
\label{AppHD20}
The full set of observational constraints used to develop the model are described in \citet{Hensley+Draine_2020b}. In brief, the extinction curve is primarily a synthesis of those of \citet{Fitzpatrick+etal_2019} in the UV and optical, \citet{Schlafly+etal_2016} in the optical and near-infrared, and \citet{Hensley+Draine_2020a} in the mid-infrared. The normalization $N_{\rm H}/E\left(B-V\right) = 8.8\times10^{21}~$cm$^{-2}$~mag$^{-1}$ is used to normalize extinction to the hydrogen column \citep{Lenz+Hensley+Dore_2017}. The infrared emission in both total intensity and polarization are based on the analyses presented in \citet{Planck_Int_XVII}, \citet{Planck_Int_XXII}, and \citet{Planck_2018_XI}, including the normalization to $N_{\rm H}$. The solid phase interstellar abundances are re-determined in \citet{Hensley+Draine_2020b} using a set of Solar abundances \citep{Asplund09, Scott+etal_2015b, Scott+etal_2015a}, a measurement of Galactic chemical enrichment from Solar twin studies \citep{Bedell+etal_2018}, and determination of the gas phase abundances from absorption spectroscopy \citep{Jenkins09}. $M_{\rm dust}/M_{\rm H} = 1.0\times 10^{-2}$ in this model.
\citetalias{Hensley+Draine_2020c} assumes the same radiation field as the \citetalias[][]{DL07} model to heat the dust grains, with updates from \citet[][]{DraineBook}.

\renewcommand{\arraystretch}{1.1}
\movetabledown=2.2in
\begin{rotatetable*}
\begin{deluxetable*}{lllll}
\tablecaption{Physical Models Calibration Summary}
    \centering
    \tablehead{
    \colhead{} & \colhead{\citet[][]{DL07}} & \colhead{\citet[][]{Compiegne11}} & \colhead{\citetalias[][]{THEMIS}} & \colhead{\citet[][]{Hensley+Draine_2020c}} }
    \startdata
    \hline
    \hline
    Extinction curve & \citet[][]{Fitzpatrick99} & \citet[][]{Mathis90} & \citet[][]{Mathis90} & \citet{Fitzpatrick+etal_2019} \\
     & & & & \citet{Schlafly+etal_2016} \\
     & & & & \citet{Hensley+Draine_2020a} \\
    $N_{\rm H}/E(B-V)$ & $5.8 \times 10^{21}$ H/cm$^2$ & $5.8 \times 10^{21}$ H/cm$^2$ & $5.8 \times 10^{21}$ H/cm$^2$ \tablenotemark{a} & $8.8\times10^{21}$~cm$^{-2}$~mag$^{-1}$ \\
    Emission spectrum & \citet[][]{Onaka96} & compiled in \citet[][]{Compiegne11} & compiled in \citet[][]{Compiegne11} & \citet{Planck_Int_XVII} \\
     & \citet[][]{Tanaka96} & & & \citet{Planck_Int_XXII} \\
     & \citet[][]{Arendt98} & & & \citet{Planck_2018_XI} \\
     & \citet[][]{Finkbeiner99} & & & \\
    $M_{\rm d}/M_{\rm H}$ & $1.0\times10^{-2}$ & $1.02\times 10^{-2}$ & $7.4\times 10^{-3}$ & $1.0\times10^{-2}$ \\
    Radiation Field & \citet[][]{Mathis83} & \citet[][]{Mathis83} & \citet[][]{Mathis83} & \citet[][]{DraineBook} \\
    \hline
    \multicolumn{5}{c}{Renormalization: emission constraint \emph{only} from \citet[][]{Compiegne11}\tablenotemark{b} and forcing $M_{\rm d}/M_{\rm H}=6.6\times10^{-3}$} \\
    \hline
    $U_{\rm min}^{\rm MW}$ & 0.6 & 1.0 & 1.0 & 1.6\\
    Normalization factor & 3.1 & 2.1 & 1.5 & 2.5 \\ 
    \enddata
\tablenotetext{a}{additional constraint: $\tau_{250}/E(B-V)=5.8\times 10^{-4}~$ \citep[][]{PlanckCollabTHEMIS}}
\tablenotetext{b}{Corrected for molecular gas only, not ionized gas.}
\tablecomments{All models share $R_V=3.1$.}
\label{TabModelRecap}
\end{deluxetable*}
\end{rotatetable*}

\section{Fitted parameter maps}
\label{AppPrms}
\begin{figure*}
    \centering
    \includegraphics[width=\textwidth, trim={5cm 12cm 5cm 3cm}, clip]{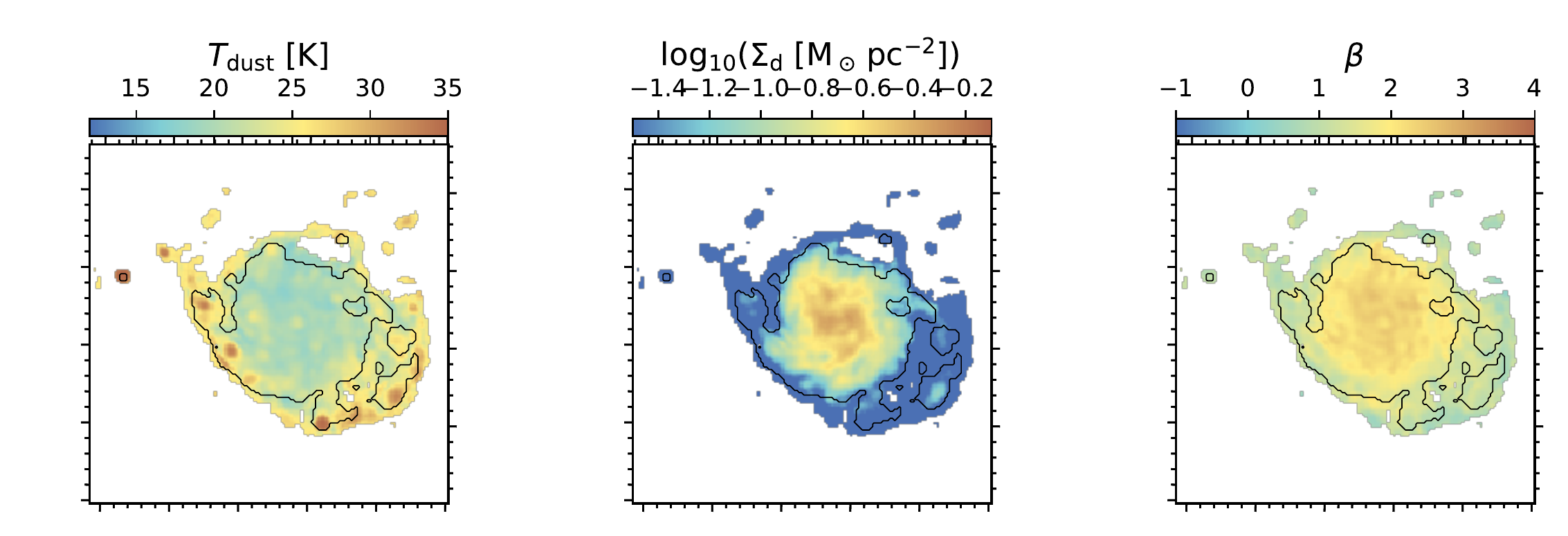}
    \hfill
    \includegraphics[width=\textwidth, trim={5cm 12cm 5cm 3cm}, clip]{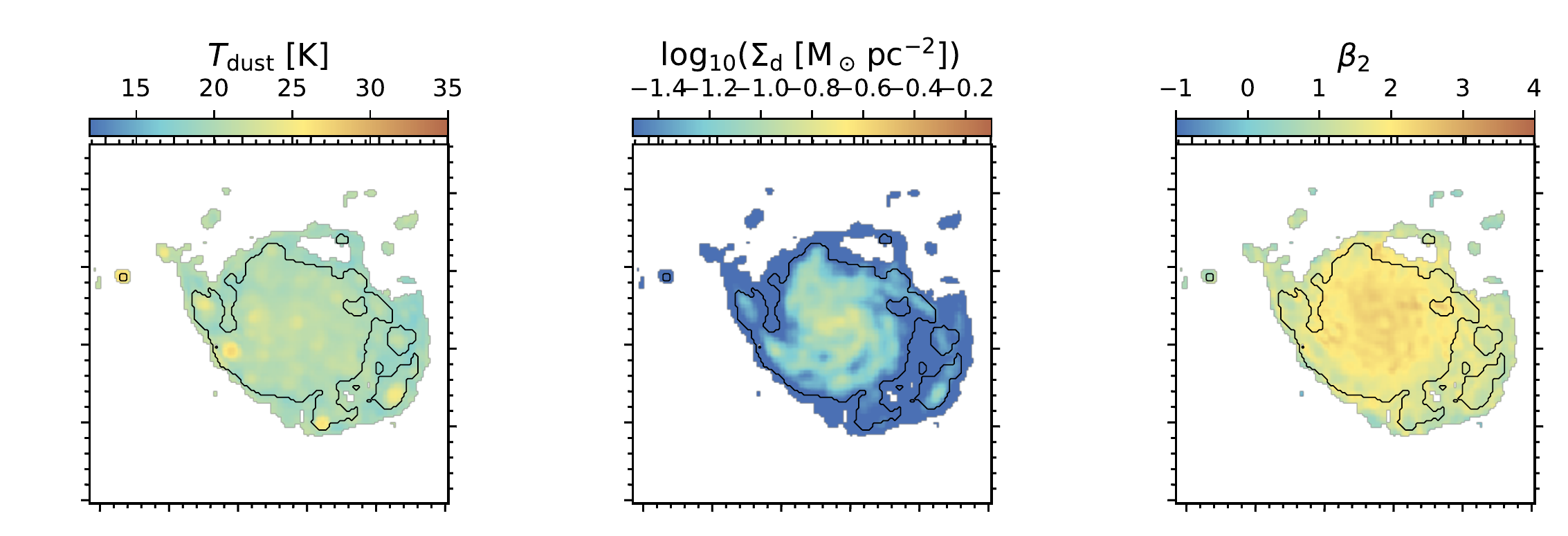}
    \caption{Maps of fitted parameters. {\it Top:} \SE  model, dust temperature (\td), total dust surface density (\sigd) and spectral index ($\beta$). {\it Bottom:} \BE  model, dust temperature (\td), total dust surface density (\sigd) and second spectral index ($\beta_2$); the breaking wavelength is fixed ($\lambda_{\rm c} = 300~\mu$m) as well as the first spectral index ($\beta=2$).}
    \label{AppPrmsBB}
\end{figure*}

\begin{figure*}
    \centering
    \includegraphics[width=\textwidth, trim={5cm 4cm 5cm 3cm}, clip]{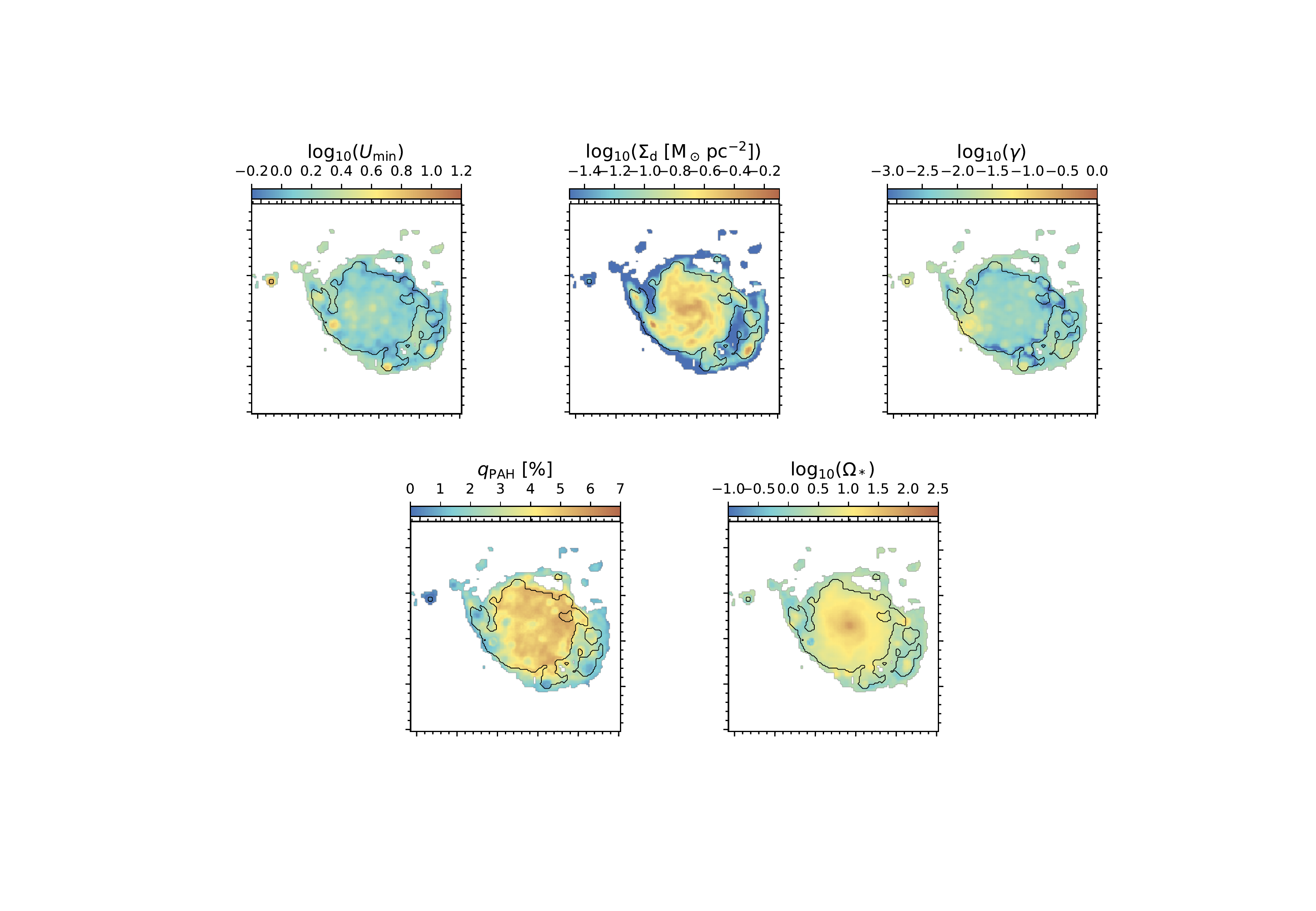}
    \caption{Maps of the fitted parameters for the \citet[][]{DL07} model: minimum radiation field (\umin), total dust surface density (\sigd), fraction of dust mass heated by a power-law distribution of radiation field ($\gamma$), PAH fraction (mass in grains with less than $10^3$ C atoms, \qpah), and scaling parameter of surface brightness (5,000~K blackbody, $\Omega_*$).}
    \label{AppPrmsDL}
\end{figure*}

\begin{figure*}
    \centering
    \includegraphics[width=\textwidth, trim={5cm 4cm 5cm 3cm}, clip]{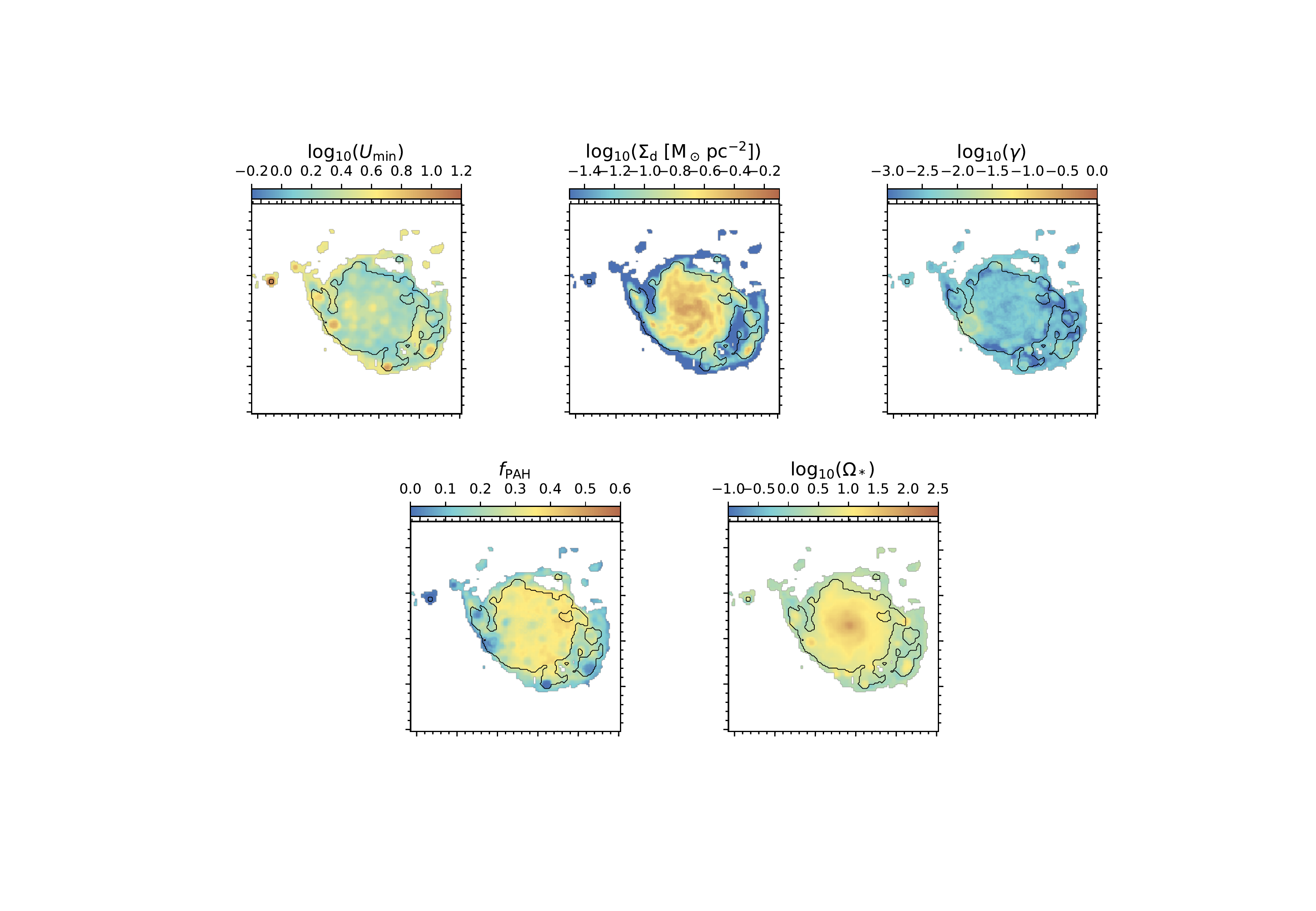}
    \caption{Maps of the fitted parameters for the \citet[][]{Compiegne11} model: minimum radiation field (\umin), total dust surface density (\sigd), fraction of dust mass heated by a power-law distribution of radiation field ($\gamma$), PAH fraction (with respect to total dust mass, $f_{\rm PAH}$), and scaling parameter of surface brightness (5,000~K blackbody, $\Omega_*$).}
    \label{AppPrmsMC}
\end{figure*}

\begin{figure*}
    \centering
    \includegraphics[width=\textwidth, trim={5cm 4cm 5cm 3cm}, clip]{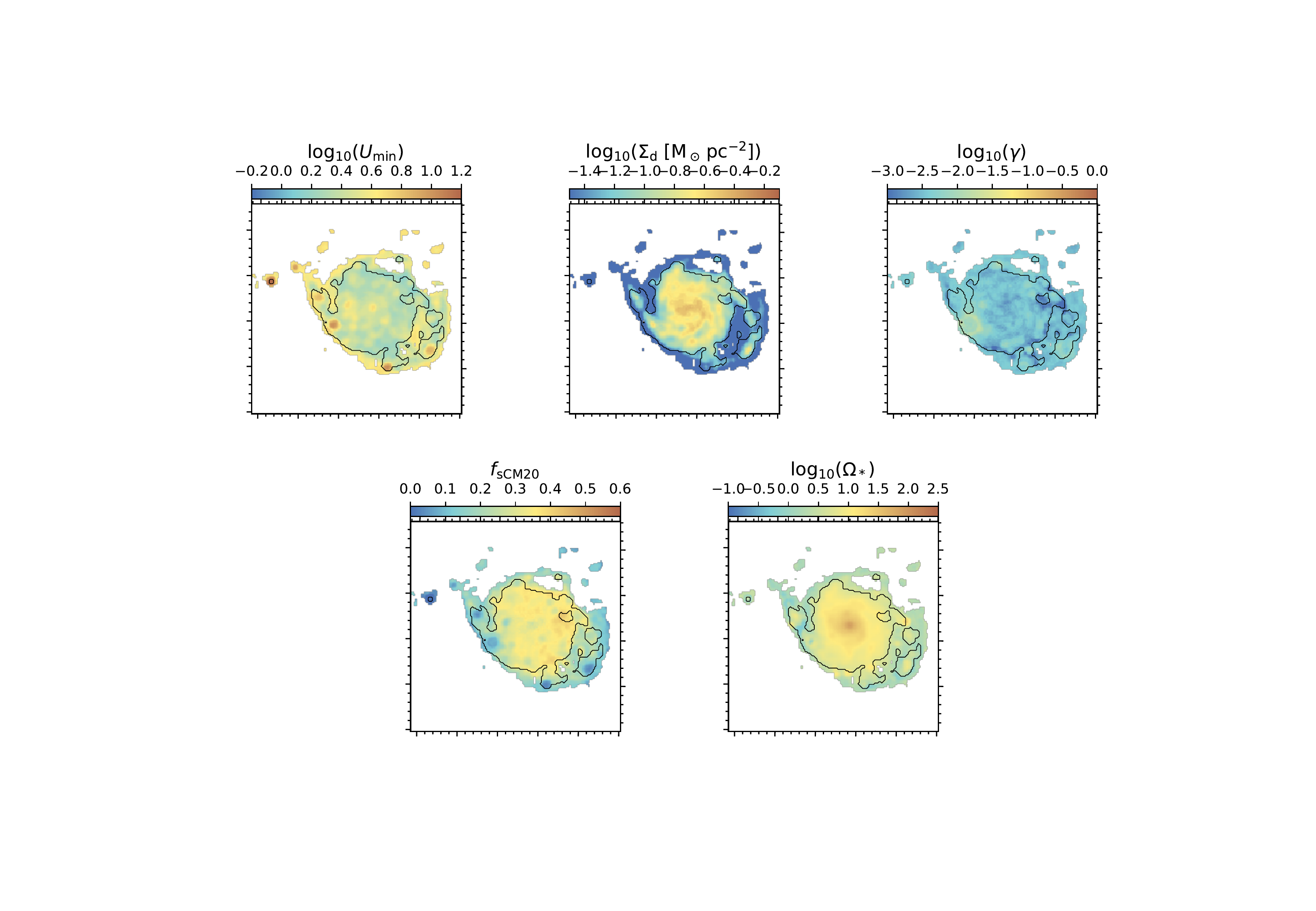}
    \caption{Maps of the fitted parameters for the \citet[][]{Jones13} model: minimum radiation field (\umin), total dust surface density (\sigd), fraction of dust mass heated by a power-law distribution of radiation field ($\gamma$), sCM20 fraction (small carbon grains, $f_{\rm sCM20}$), and scaling parameter of surface brightness (5,000~K blackbody, $\Omega_*$).}
    \label{AppPrmsTHEMIS}
\end{figure*}

\begin{figure*}
    \centering
    \includegraphics[width=\textwidth, trim={5cm 4cm 5cm 3cm}, clip]{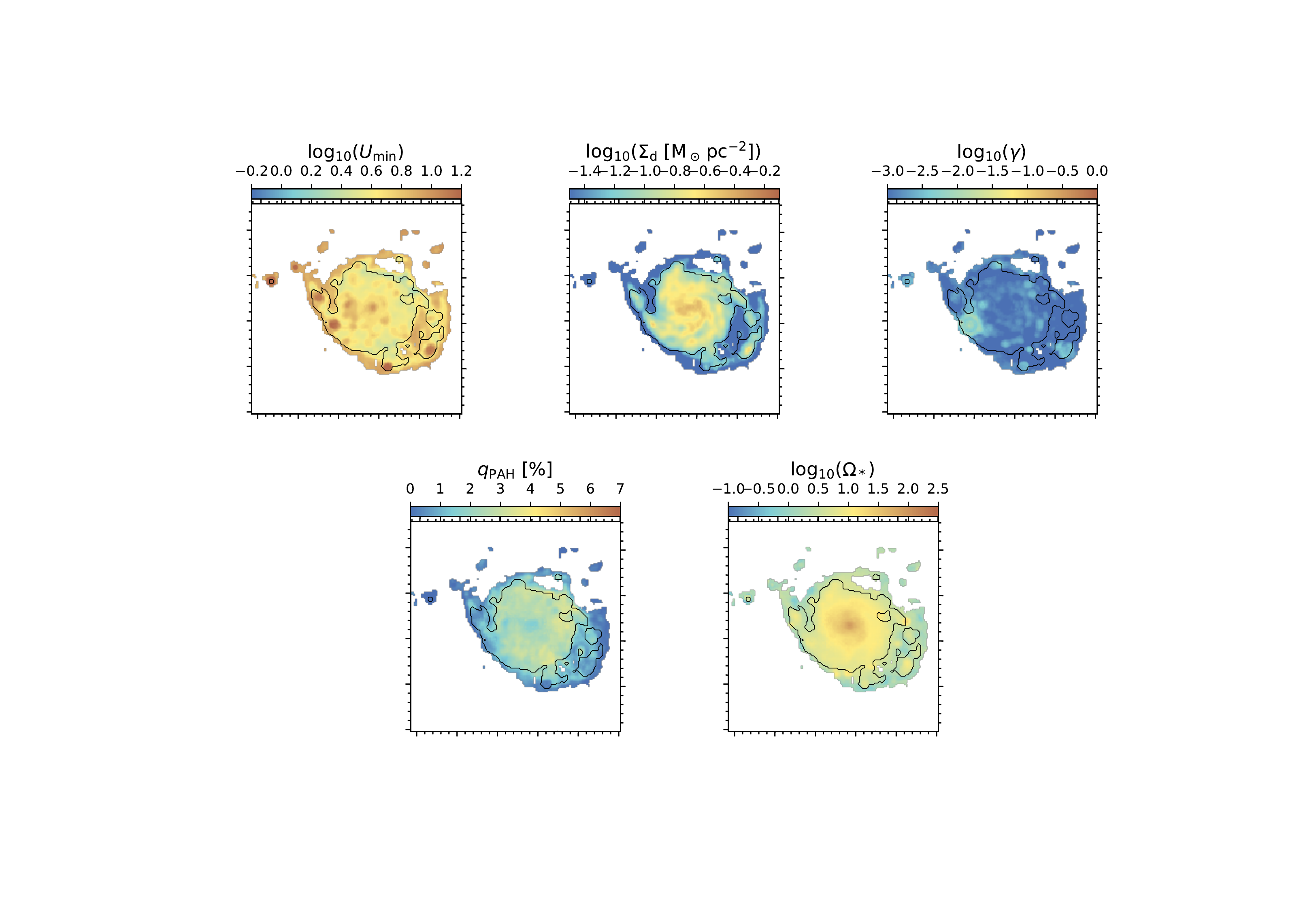}
    \caption{Maps of the fitted parameters for the \citet{Hensley+Draine_2020c} model: minimum radiation field (\umin), total dust surface density (\sigd), fraction of dust mass heated by a power-law distribution of radiation field ($\gamma$), PAH fraction (mass in grains with less than $10^3$ C atoms, \qpah), and scaling parameter of surface brightness (5,000~K blackbody, $\Omega_*$).}
    \label{AppPrmsHD}
\end{figure*}

We present the spatial variations of the fitted parameters for all six models, in Figures~\ref{AppPrmsBB} -- \ref{AppPrmsHD}. The contours mark the 3$\sigma$ detection threshold. 
We use the same scale for identical parameters, when possible.


\section{Residual maps}
\label{AppResd}
\begin{figure*}
    \centering
    \includegraphics[width=1.3\textwidth, angle =90]{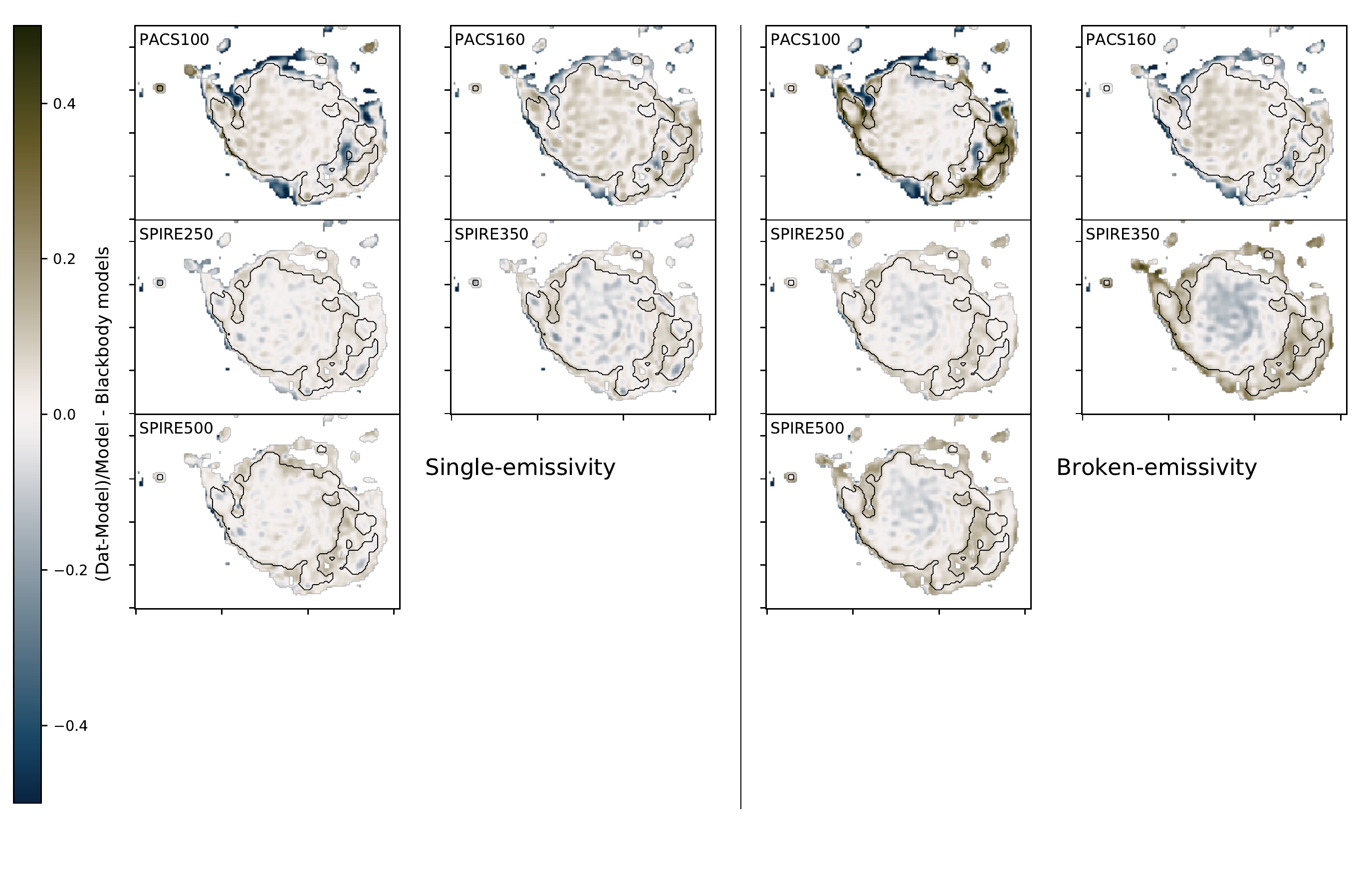}
    \caption{Maps of the fractional residuals for the \SE and \BE models.}
    \label{AppResdBB}
\end{figure*}

\begin{figure*}
    \centering
    \includegraphics[width=1.3\textwidth, angle =90]{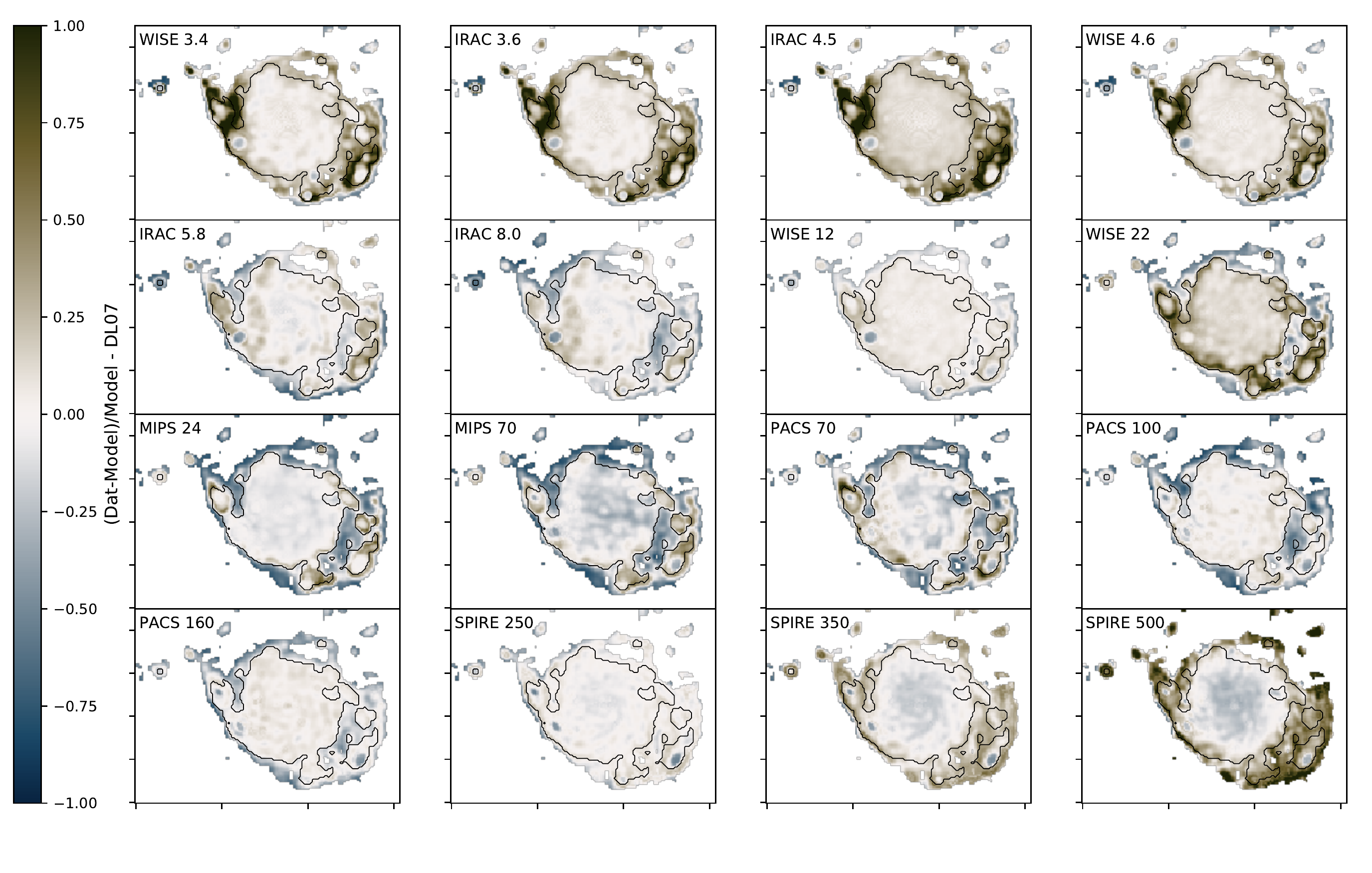}
    \caption{Maps of the fractional residuals for the \citet[][]{DL07} model.}
    \label{AppResdDL}
\end{figure*}

\begin{figure*}
    \centering
    \includegraphics[width=1.3\textwidth, angle =90]{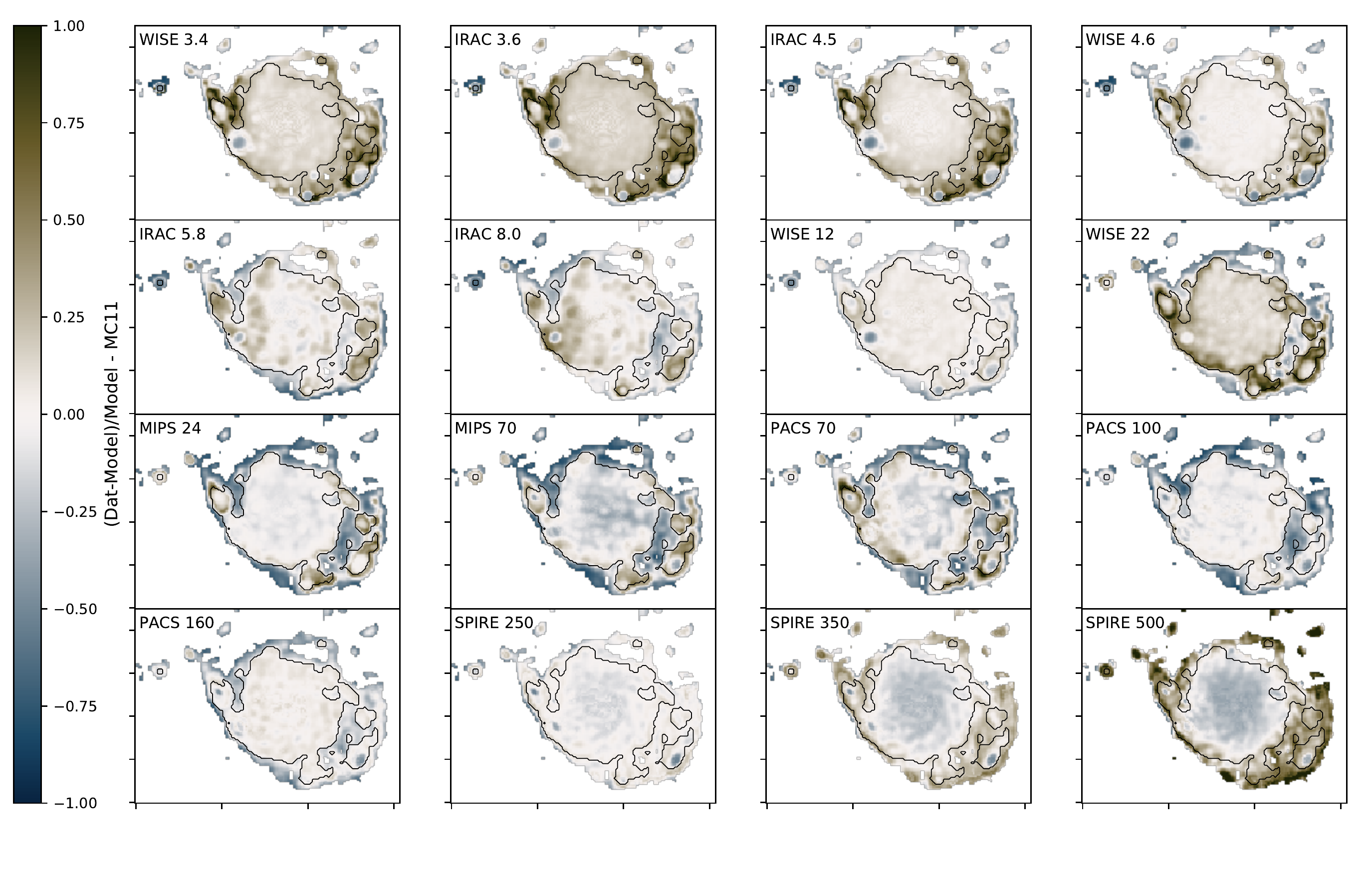}
    \caption{Maps of the fractional residuals for the \citet[][]{Compiegne11} model.}
    \label{AppResdMC}
\end{figure*}

\begin{figure*}
    \centering
    \includegraphics[width=1.3\textwidth, angle =90]{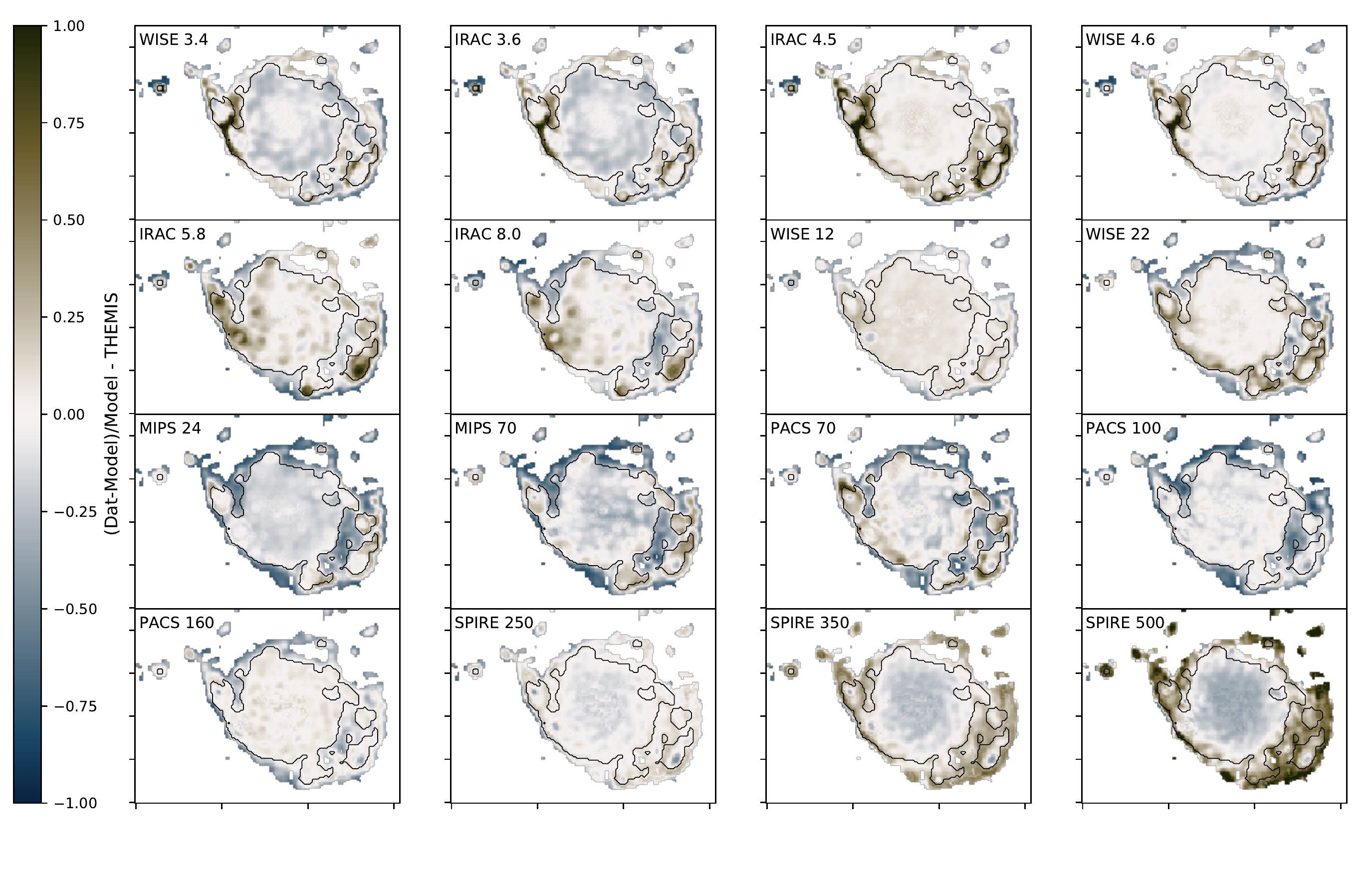}
    \caption{Maps of the fractional residuals for the \citet[][]{Jones13} model.}
    \label{AppResdTHEMIS}
\end{figure*}

\begin{figure*}
    \centering
    \includegraphics[width=1.3\textwidth, angle =90]{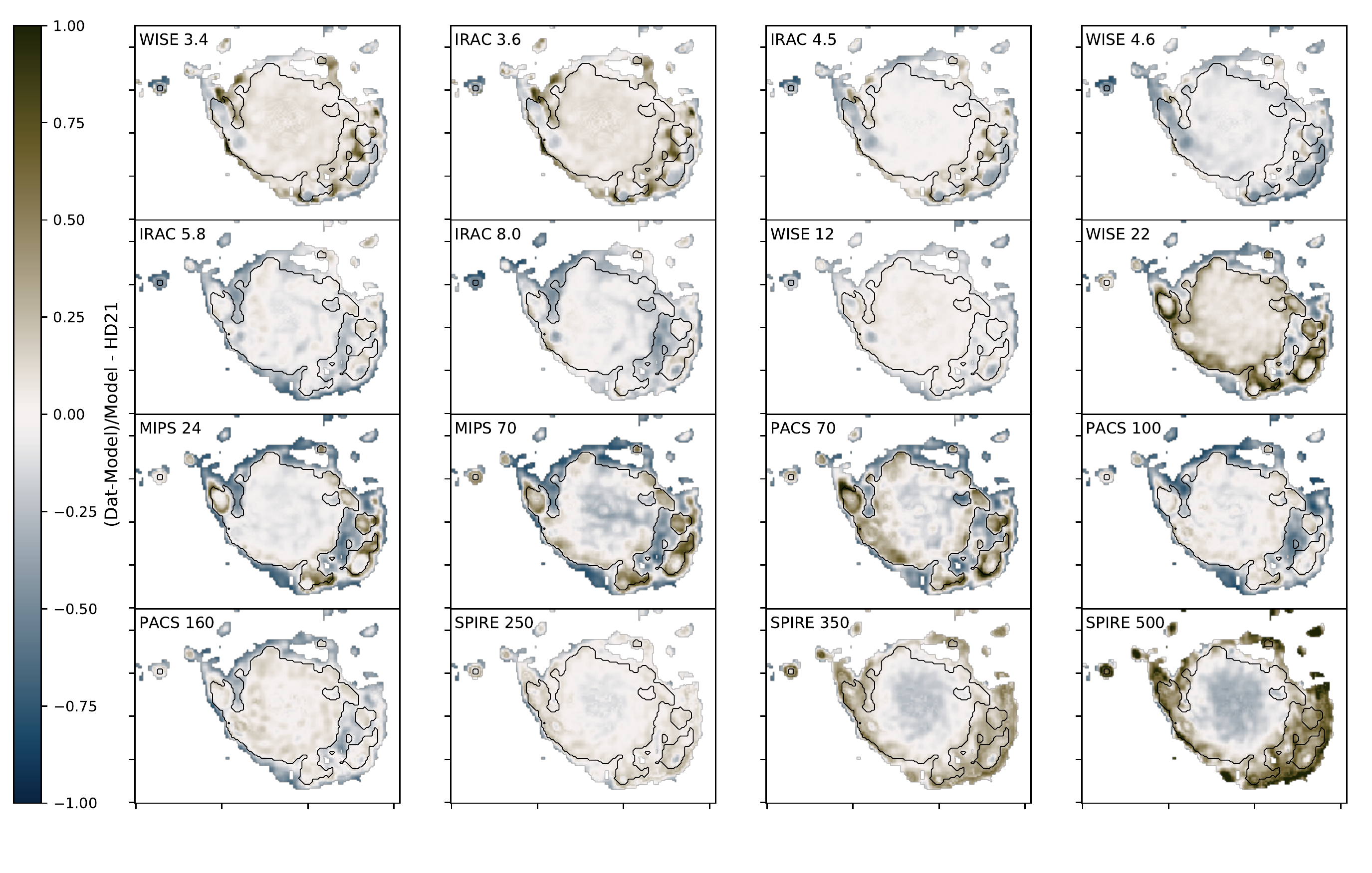}
    \caption{Maps of the fractional residuals for the \citet{Hensley+Draine_2020c} model.}
    \label{AppResdHD}
\end{figure*}

We present the spatial variations of the fractional residuals \textit{(Data-Model)/Model} for all six models, in Figures~\ref{AppResdBB} -- \ref{AppResdHD}. The contours mark the 3$\sigma$ detection threshold. 
The so-called ``sub-millimeter'' excess is visible in most maps at SPIRE~500 (blue shade).


\section{Fits quality}
\label{AppMaxLkd}
We show the relative quality of the fits between each model. The value displayed in the maximum likelihood, in arbitrary units. The \SE and \BE models show the least dynamic range but never reach the highest values of the physical models. For the physical models, we can clearly see the \ion{H}{2} regions showing fits with low confidence, likely related to the issues mentioned in Section~\ref{SecSpireSigd}.

\begin{figure*}
    \centering
    \includegraphics[width=\textwidth, trim={4cm 2.5cm 1.5cm 2.5cm}, clip]{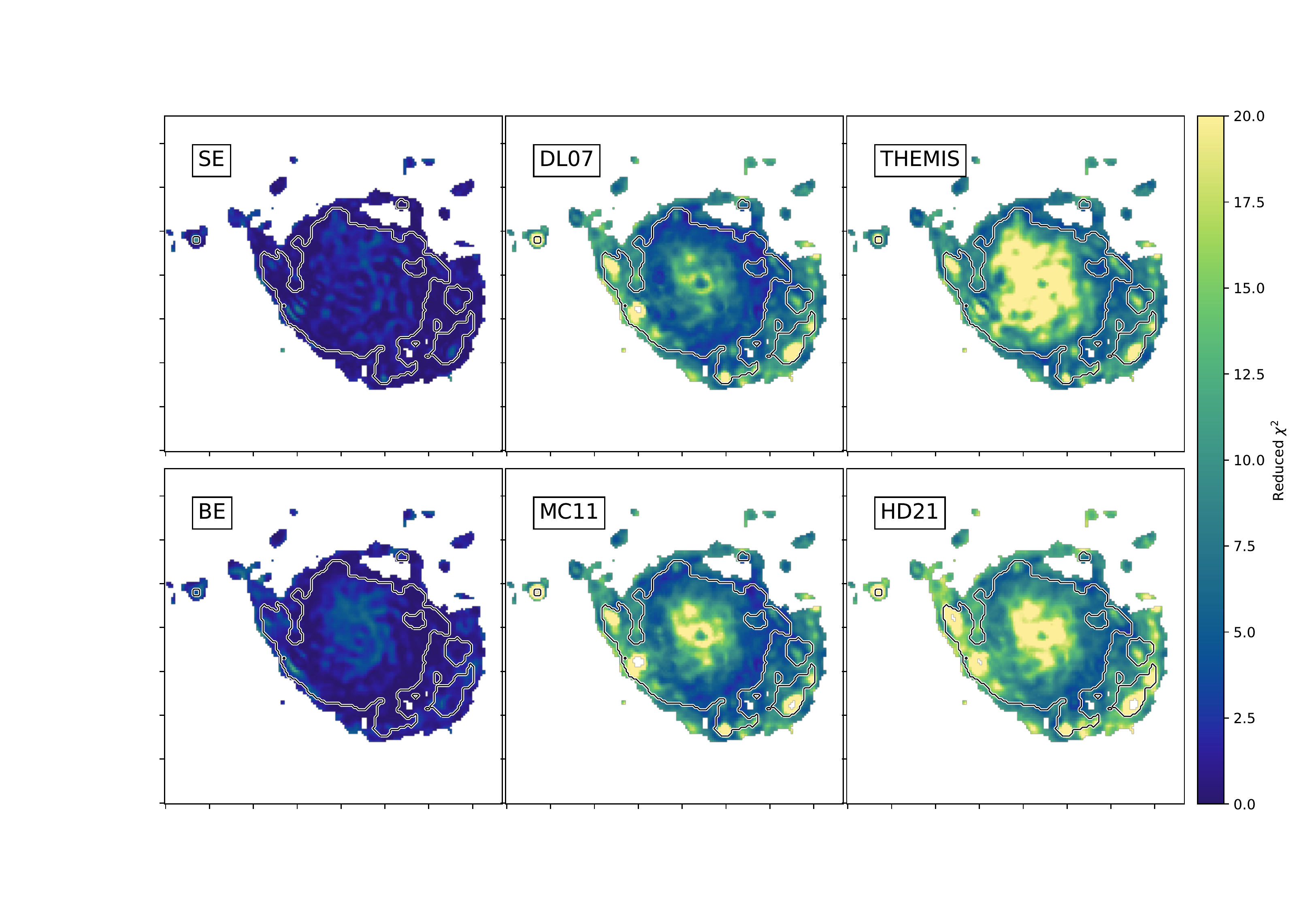}
    \caption{Maps of the reduced $\chi^2$ in each pixel. The \ion{H}{2} regions show the lowest confidence for the physical models, while the modified blackbodies show good fits on the entire disk. The contour marks the 3-$\sigma$ detection threshold.}
    \label{FigMaxLkd}
\end{figure*}



\end{document}